\documentclass[a4paper,fleqn,usenatbib,dvipdf]{mnras}

\usepackage{newtxtext,newtxmath}

\usepackage[T1]{fontenc}
\usepackage{ae,aecompl}

\usepackage{graphicx}	
\usepackage{amsmath}	
\usepackage{amssymb}	
\usepackage{bm}
\usepackage{subfigure}
\usepackage{url}
\usepackage{color}

\title[Environmental effects around underdensities]{Environmental effects on halo abundance and weak lensing peak statistics toward large underdense regions}

\author[Y.Higuchi and K.T.Inoue]{
Yuichi Higuchi,$^{1, 2, 3}$\thanks{E-mail: yuichi.higuchi@nao.ac.jp}
Kaiki Taro Inoue$^{2}$
\\
$^{1}$National Astronomical Observatory of Japan (NAOJ), 2-21-1 Osawa, Mitaka, Tokyo 181-8588, Japan\\
$^{2}$Faculty of Science and Engineering, Kindai University, Higashi-Osaka, Osaka, 577-8502, Japan\\
$^{3}$Academia Sinica Institute of Astronomy and Astrophysics (ASIAA), No. 1, Section 4, Roosevelt Rd, Taipei 10617, Taiwan\\
}

\date{Accepted XXX. Received YYY; in original form 2018}

\pubyear{2018}

\begin{document}
\label{firstpage}
\pagerange{\pageref{firstpage}--\pageref{lastpage}}
\maketitle

\begin{abstract}
The cosmic microwave background (CMB) contains an anomalous cold spot with a surrounding hot ring, known as the Cold Spot.
 \citet{2006ApJ...648...23I} proposed that this feature could be explained by postulating a supervoid:
if such a large underdense region exists, then the growth of matter perturbing around the spot might differ from the average value in the Universe and the differences might affect weak lensing analysis of peak statistics. 
To investigate environmental effects on halo number count and peak statistics, we used a publicly available ray-tracing simulation for a box size of 2250$h^{-1}$Mpc on a side \citep{2017ApJ...850...24T}. 
We found that the number counts for massive haloes toward the largest underdense region in the simulation decreases and the corresponding significance of the difference, based on a cosmic average, is $\geq3\sigma$.
On the basis of the results of peak statistics analysis, the number of high peaks decreases with the decrement of massive haloes, but the number of low peaks increases with the lack of matter in the line of sight.
The highest significance of the decrement in peak counts in large underdense regions is $5\sigma$ in the total signal-to-noise ratio. 
Our result implies that environmental effects on halo abundance and weak lensing peak statistic can be used to probe the presence and properties of supervoids.
\end{abstract}




\section{Introduction}
Although results of cosmic microwave background (CMB) experiments and large-scale surveys are well consistent with predictions from the $\Lambda$CDM model \citep{2004PhRvD..69j3501T, 2005ApJ...633..560E, 2013ApJS..208...19H, 2015arXiv150201589P}, 
some anomalous features in the CMB temperature map, such as the Cold Spot \citep{2004ApJ...609...22V, 2005MNRAS.356...29C,2013ApJS..208...19H, 2014A&A...566A.135G,2016A&A...594A..16P}, have been observed. 
Such features have been explained by \citet{2006ApJ...648...23I} who proposed supervoids with a radius of $200-300h^{-1}$Mpc in local Universe. 
Although there is a temperature decrease of approximately $20$ $\mu$K due to the potential decay of a supervoid, the integrated Sachs-Wolfe effect seems too small but sufficient to explain the Cold Spot: the remaining contribution can be explained by the ordinary Sachs-Wolfe effect at the last scattering surface if there is a positive density perturbation in the line of sight to the Cold Spot \citep{IST10, 2012MNRAS.421.2731I}.     
In addition to the CMB observations, some photometric and spectroscopic observations of galaxies revealed the presence of such a large underdense region with some overdense structures toward the Cold Spot in the galaxy distribution \citep{2010ApJ...714..825G, 2016MNRAS.455.1246F, 2016MNRAS.462.1882K, 2017MNRAS.470.2328M}; 
however, its size and redshift properties have not been well understood because of large uncertainty in the photometric redshift and time limitation for spectroscopic observations.

Moreover, the growth of haloes toward large underdensities has been an unclear subject in terms of investigation of environmental effects on the evolution of haloes, which is an important aspect of their growth history.
Analytical studies showed that the clustering amplitude of haloes depends only on halo mass \citep{1984ApJ...284L...9K, 1996MNRAS.282..347M, 2002PhR...372....1C},
whereas studies of {\it N}-body simulations argued that amplitude and halo mass functions could not be fully described by simple analytical formulae
as these properties depend on other environmental parameters such as concentration, formation epoch and spin of the host halo, called the assembly bias \citep{2002PhR...372....1C, 2007ApJ...657..664J, 2008ApJ...688..709T, 2009MNRAS.398.1742H, 2010ApJ...712..484F, 2010ApJ...708..469F, 2018MNRAS.474.5143M}. 
The observational evidence remains under discussion.
Some observations obtained through weak gravitational lensing signals and correlation functions \citep{2014MNRAS.443.3107L, 2016ApJ...819..119L, 2016PhRvL.116d1301M, 2017MNRAS.470..551Z, 2017MNRAS.470.4767B}, from the {\it Sloan Digital Sky Survey} \citep{2000AJ....120.1579Y}, provided hints of the assembly bias where large underdense environment should affect halo growth inside the region that might be detected with on-going and future large-scale surveys as the {\it Subaru/Hyper Suprime-Cam (HSC)} \citep{2006SPIE.6269E...9M} and {\it Dark Energy Survey (DES)} \citep{2012AAS...21941305D}.

Furthermore, peak statistics in weak gravitational lensing, which counts the number of peaks in a convergence field, has been proposed to extract cosmological information and was adopted to the observational data \citep{2010A&A...519A..23M, 2015PASJ...67...34H, 2016MNRAS.459.2762H, 2016MNRAS.463.3653K, 2017MNRAS.466.2402S}.
As weak lensing could trace matter distribution on a light path and a change in the halo mass function could influence the number of weak lensing peak counts \citep{2004MNRAS.350..893H}, 
the presence of a surpervoid could be confirmed and properties associated with weak lensing peak counts and halo number counts could be investigated given environmental effects that are significant on the matter growth.  

On-going photometric and spectroscopic large-scale surveys enable to identify a large number of galaxies and carry out weak lensing analysis toward the supervoids.
Analysis of galaxy distribution and weak lensing  with such deep and wide surveys can reveal the presence of the supervoids if their presence affect the growth of matter.
In order to confirm the presence with the surveys, we investigate the environmental effects in large underdense regions on halo abundance and weak lensing peak counts through full-sky ray-tracing simulations of low convergence regions generated on convergence maps by underdense regions with some overdense regions, assuming a concordant $\Lambda$CDM model.
Although the observed supervoid contains interior structures (e.g., haloes and filaments), the density contrast in the region as a whole becomes negative.
The structure of the paper is organised as follows. 
Section~\ref{sec.ana} describes the basics of weak lensing and the methods of analysis.
Section~\ref{sec.sim} briefly introduces the simulation and the corresponding search algorithm for the large underdense regions in the simulation. 
Section~\ref{sec.result} presents the result of the  halo mass function and peak statistics. 
Lastly, Section~\ref{sec.con} provides a summary of all results . 
The cosmological parameters employed through this paper follow the {\it WMAP~9yr} result \citep{2013ApJS..208...19H}: 
the Hubble parameter $H_0=70$ km/s/Mpc, density parameter of total
matter $\Omega_{\rm m}=0.279$, $\Omega_\Lambda=0.721$, spectral index
$n_s=0.972$ and density fluctuation amplitude $\sigma_8=0.823$, assuming a flat FLRW cosmology.

\section{Data analysis}
\label{sec.ana}
\subsection{Principles of weak lensing}
Gravitational lensing traces matter distribution between a source galaxy and an observer, at an effect characterized through a convergence $\kappa$ and a shear $\gamma$.    
The Convergence can be expressed in terms of the lensing weight function $W(\chi)$ and the density contrast $\delta\left( \chi\theta, \chi \right)$ as \citep{2001PhR...340..291B}
\begin{equation}
\kappa\left(\bm{\theta}\right)=\int^{\infty}_{0} {\rm d}\chi W\left(\chi\right) \delta\left(\chi\bm{\theta},\chi\right),
\label{eq.con}
\end{equation}
where $\chi$ is a comoving distance.
The lensing efficiency depends on the redshifts of source galaxies and lenses, whereas the lensing weight function
\begin{equation}
W\left(\chi\right)=\frac{3H_0^2\Omega_{\rm m0}}{2c^2}q(\chi)\left\{1+z\left(\chi\right)\right\},
\end{equation}
where $q(\chi)$ is defined as
\begin{equation}
q(\chi)=S_K(\chi) \int^{\infty}_{\chi}{\rm d}\chi' w_{\rm s}\left(\chi'\right) \frac{S_K(\chi'-\chi)}{S_K(\chi')}.
\end{equation}
$w_{\rm s}\left(\chi\right)$ indicates the number distribution of source galaxies in a given line of sight.
Assuming a flat FLRW model, $S_k(\chi)$ becomes
\begin{equation}
S_K(\chi)=\chi
\end{equation}
 
\subsection{Peak statistics} 
A convergence either higher or lower at a pixel than at surrounding pixels is defined as a weak lensing "peak"  \citep{2004MNRAS.350..893H}.
Some high peaks in a peak distribution correspond to haloes \citep{2004MNRAS.350..893H, 2016MNRAS.459.2762H}.
Peak values are defined as a ratio between a smoothed convergence value and a noise value as 
\begin{equation}
\nu = \frac{\kappa_s}{\sigma_{\rm noise}},
\label{eq.sn_peak}
\end{equation} 
where a smoothed convergence $\kappa_s$ is obtained by convolving with a filter $W_F(x)$ in
\begin{equation}
\kappa_s\left(\theta_s\right) = \int{\rm d}^2\phi W_F(\phi; \theta_s) \kappa(\phi),
\end{equation}
and $\theta_s$ is a smoothing scale, smoothed by a Gaussian filter
\begin{equation}
W_F(x) = \frac{1}{\pi\theta_s^2}{\rm exp}\left(-\frac{x^2}{\theta_s^2}\right).
\end{equation}
If we assume that a noise follows a Gaussian random field, the variance of a noise is calculated by \citep{2000MNRAS.313..524V}
\begin{equation}
\sigma_{\rm noise}^2 = \frac{\sigma_{\rm e}^2}{2}\frac{1}{2\pi n_g \theta_s^2},
\end{equation}
where  $\sigma_{\rm e}$ is the root mean square of amplitudes of galaxy ellipticities used in a survey and $n_g$ is the background galaxy number density.
On small scales, we need to take into account the contribution from clustering of background galaxies \citep{2000MNRAS.313..524V}. 
\citet{1998ApJ...499L.125C} showed that the angular correlation length is a few arcsec.
Since we conducted a degree scale study, we ignored this effect.

In this paper, we assumed $n_g = 20$ arcmin$^{-2}$ and $\sigma_{\rm e}=0.4$, respectively, which are adopted from the values in the {\it HSC} and {\it DES} surveys \citep{2018PASJ...70S..25M, 2018MNRAS.481.1149Z}. 
To observe their effects on peak statistics around local supervoids with these survey data, we fixed the source redshift $z_s=0.713$ as in the {\it DES} survey.

\subsection{Stacking analysis}
Stacking helps averaging out observational noise and characteristic properties of samples.
In order to investigate their average properties and compare with theoretical models, the stacked lensing method has been applied \citep{2010PASJ...62..811O, 2011PhRvD..83b3008O, 2013MNRAS.432.1021H, 2016ApJ...821..116U, 2017ApJ...836..231U, 2018MNRAS.478.4277S}.
In our analysis, we investigated the average effects on halo number count and peak statistics in underdense regions by stacking the results obtained in each underdense region. 
The average values for the halo number count/peak count $<d_i>$ and the dispersion $\sigma_i$ in $i$-th bin for $N$-samples are obtained by
\begin{equation}
<d_i> = \sum_{j=1}^{N}\frac{d_{i, j}}{N},
\end{equation}
\begin{equation}
\sigma_i^2 = \sum_{j=1}^{N} \frac{(d_{i, j} - <d_i>)^2}{N},
\end{equation}
where $d_{i, j}$ is the value in $i$-th bin for $j$-th sample. 
With the stacking method, we will show the average effects in underdese regions.

\subsection{Halo size estimation on sky plane}
The halo density profile is described well by the Navarro-Frenk-White (NFW) profile \citep{1997ApJ...490..493N},
\begin{equation}
\rho\left( r \right) = \frac{\delta_c\rho_c}{r/r_s\left(1+r/r_s\right)^2},
\end{equation}
where $\rho_c$ and $r_s$ are the critical density and the scale radius, respectively.
$\delta_c$ is a characteristic overdensity defined with a concentration parameter $c=r_{\rm 200}/r_s $ by
\begin{equation}
\delta_c = \frac{\Delta_{200}}{3}\frac{c^3}{\textrm{ln}(1+c)-c/(1+c)}.
\end{equation}
$M_{200}$ is defined such that the average density of a halo within a virial radius is equal to two hundred times the mean matter density of the Universe \citep{1997PThPh..97...49N}, as expressed in 
\begin{equation}
M_{\rm 200} = \frac{4\pi}{3}\Delta_{\rm 200}\left(z\right) \bar{\rho}\left(z\right)r^3_{\rm 200},
\end{equation}
\begin{equation}
c_{\rm 200} = \frac{1}{r_s}\left[ \frac{3M_{\rm 200}}{4\pi\Delta_{\rm 200}\left(z\right)\bar{\rho}\left(z\right)} \right]^{1/3}.
\end{equation}
The concentration parameter correlates with masses and redshifts of haloes, and is obtained from $N$-body simulations \citep{2008MNRAS.391.1940M} such that 
\begin{equation}
c_{\rm vir}\left(M_{\rm vir}, z\right) = 7.26\left(\frac{M_{\rm vir}}{10^{12}h^{-1}{\rm M}_\odot} \right)^{-0.086} \left(1+z\right)^{-0.71}.
\label{eq.concent}
\end{equation}
Most of high peaks in the peak statistics correspond to massive haloes.
By adopting a smoothing filter to a map with a smoothing scale, smaller scale structures are smoothed out and structures whose sizes are similar to the smoothing scale can be picked up.
To maximize the difference in the peak statistics, we smoothed a convergence map with a typical size of a massive halo.
By assuming a halo mass, the NFW profile and the fitting formulae in Section~\ref{sec.sim},  we estimate the smoothing scale and smoothed the maps with it.
For more details, please see the third paragraph  in Section~\ref{sec.sim}

\subsection{Signal-to-noise ratio}

We quantify the significance of the difference between profiles for a random field and for a large underdense region, using the signal-to-noise $(S/N)$ ratio as 
\begin{equation}
\left( \frac{S}{N} \right)^2 =\sum_{i,j} \left[ N_{v, i} - N_{r, i} \right] \left[{\bm C}\right]_{ij}^{-1} \left[ N_{v, j} - N_{r, j} \right],
\label{eq.sn}
\end{equation}
where $\bm{C}$ is a covariance matrix and $\bm{C}^{-1}$ is its inverse, and $N_{v, i}$ and $N_{r, i}$ are the number counts in the $i$ bin for the underdense region and random field, respectively.
The covariance matrix is composed of the statistical uncertainty and shot noise defined by
\begin{equation}
{\bm C}={\bm C}^{\rm stat}+{\bm C}^{\rm shot}.
\end{equation}
In our analysis, the statistical uncertainty was calculated by selecting 500 random points from the 108 realisations and measuring the variance of the number count.
The shot noise was estimated by assuming a Poisson error.
For the stacking analysis, we replace $N_{v, i}$ with $\langle N_{v, i} \rangle$, which is an average value in the low underdense regions.

\subsection{Press-Schechter theory}
The Press-Shcechter theory provides a good prediction of the number of haloes in a given simulation \citep{1974ApJ...187..425P}.
The density contrast can be described by the Gaussian probability distribution function
\begin{equation}
P(\delta) = \frac{1}{\sqrt{2\pi\sigma^2\left(M\right)}} {\rm exp}\left(-\frac{\delta^2}{2\sigma^2\left(M\right)}\right),
\end{equation}
where $\sigma\left(M\right)$ is a dispersion in the scale of mass $M$.
The halo regions collapse when fluctuations exceed a threshold $\delta_c$ predicted by a spherical collapse model.
The probability in $>M$ haloes is obtained through
\begin{equation}
P_{>\delta_c}(M) = \int^{\infty}_{\delta_c}d\delta P(\delta)=\frac{1}{2}{\rm erfc}\left(\frac{\nu}{\sqrt{2}}\right),
\end{equation}
where $\nu=\delta_c/\sigma\left(M\right)$.
Consequently, the number of haloes $n(M)dM$ between $M$ and $M+dM$ can be expressed as
\begin{equation}
n(M)dM = \sqrt{\frac{2}{\pi}} \frac{\bar{\rho}_0}{M^2}\mid \frac{d{\rm ln}\sigma\left(M\right)}{d{\rm ln}M}\mid \frac{\delta_c}{\sigma\left(M\right)} {\rm exp}\left(-\frac{\delta_c^2}{2\sigma^2\left(M\right)}\right)dM,
\label{eq.massfunc}
\end{equation}
where $\bar{\rho}_0$ is the average matter density.
The variance of the density field depends on redshift and masses of haloes as
\begin{equation}
\sigma^2(M, z) = \frac{1}{2\pi^2}\int_0^\infty dk k^2P(k,z)|\tilde{W}(kR)|^2,
\label{eq.sigmam}
\end{equation}
where $R$ is a radius for mass $M$, $P(k, z)$ is the power spectrum at redshift of $z$ with the scale of $k$ and $\tilde{W}(kR)$ is the Fourier transform of the filter function.
The transfer function is calculated with the fitting formulae obtained by \citep{1998ApJ...496..605E, 1999ApJ...511....5E} and we used the top-hat filter.
See appendix~\ref{ap:trans} for the details on the transfer function.
 
To obtain the number count of haloes inside a void, we modify the matter density parameter such that
\begin{equation}
\Omega_{\rm m}' = \Omega_{\rm m}\left(1+\delta_v\right), 
\label{eq.omega_m}
\end{equation}
where $\delta_v$ is the density contrast of a void.
We adopt the parameter to equation~(\ref{eq.transfer})-(\ref{eq.betanode}) and calculated the power spectrum with the parameter.
It changed the variance of matter by suppressing the amplitude of the matter power spectrum at small scale and the number count of haloes.

\section{ray-tracing simulation and finding algorithm for underdense regions}
\label{sec.sim}
We used a publicly available weak lensing simulation\footnote{\url{http://cosmo.phys.hirosaki-u.ac.jp/takahasi/allsky_raytracing/index.html}} for investigation of the environmental effects \citep{2017ApJ...850...24T}, using $2048^3$ particles for different box sizes.
The smallest box has minimum halo mass of $4.1\times10^{10}h^{-1}$M$_\odot$, and the total number of realisations is $108$, 
distributed into halo catalogues, lens and CMB maps for a single realisation.  
They created halo catalogues, via {\it ROCKSTAR}, a phase-space friends-of-friends algorithm, and defined lensing information at each pixel with {\it HEALPix} method \citep{2005ApJ...622..759G}. 
The simulation is detailed in \citet{2017ApJ...850...24T}.

\begin{figure}
\begin{center}
\includegraphics[width=1.0\columnwidth]{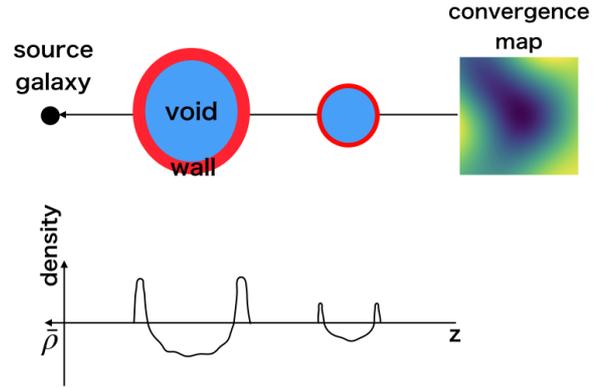}
\caption{Schematic view of the finding algorithm of underdense regions.
The upper diagram shows the positions of a source galaxy and structures in the line of sight.
The lower diagram shows the density contrast corresponding to the upper panel as a function of redshift.
Although there are some overdense regions on a light path, 
several underdense regions between a source galaxy and an observer can produce negative convergence on a convergence map.
In the simulation, underdense regions are searched by smoothed convergence maps.}
\label{fig.schview}
\end{center}
\end{figure}

Since convergence field traces a matter density between sources and an observer, contiguous underdense regions can be found by smoothing the convergence maps \citep{2018MNRAS.476..359H}.
Figure~\ref{fig.schview} shows a schematic view of our underdense search method.
Previous observations signified the presence of some underdense and overdense regions that construct a supervoid;
although some overdensities and underdensities exist between a source plane and an observer, several underdense regions in the line of sight can create a low convergence region.   
We smoothed noise-free convergence maps with a Gaussian filter and chose a smoothing scale $\theta_s=20$ degrees, which is a similar size for the Cold Spot, and a source redshift $z_s=0.713$, similar to on-going and future large-scale photometric surveys.
After the smoothing process, we searched for convergence peaks and calculated their $S/N$ according to equation~(\ref{eq.sn_peak}).
In each map, low peaks of $S/N$ values lower than $-10$ were selected and sorted in descending order. 
To exclude peaks belonging in the same underdense regions, distances between the peaks were calculated.
The peaks were rejected when their distance to higher peak was closer than the full width at half maximum $2\sqrt{2{\rm ln}2}\theta_s$.
Thus, a total of 106 large underdense regions in the 108 realisations were selected.
The pixels of the positions of the weak lensing peaks were applied as the centres of the large underdense regions.
In the subsequent analysis, we investigated environmental effects at the lowest peak and the average effects around the low peaks, respectively.
We measured the number of haloes and weak lensing peaks within a search radius $\theta_r$ on the sky plane with respect to the centres.
Moreover, we examined the average properties by stacking the calculated values measured at the 106 points. 
Although the smoothing scale affects the centres of the underdense regions, the differences are a few degrees after we use a large smoothing scale.
As this uncertainty is smaller than the search radius used in the halo and peak analysis in most cases, it does not have substantial effect on the result. 

Because the number of pixels in {\it HEALPix} is proportional to the square of N$_{\rm side}$, a high resolution map requires large computational time in carrying out the analysis of peak statistics. 
We found that the large underdense regions affect the growth of haloes and this environmental effect appeared especially in high peaks (see Section~\ref{sec.halo_pro}).
Therefore, we mainly focused on those peaks in the peak analysis. 
For a halo with a virial mass of $M_{\rm vir}=10^{15}h^{-1}{\rm M}_\odot$ for a concentration defined by equation~(\ref{eq.concent}) at redshift $z=0.4$, its angular size of this halo becomes 0.15 degree on the sky plane; 
therefore, we used slightly higher resolution maps with ${\rm N_{side}}=512$ corresponding to a resolution of 0.11 degree on the sky plane.
To mimic weak lensing observations, we added noise to the convergence maps and smoothed them with a smoothing scale of 0.15 degree.
The maps were used later for the analysis of peak statistics.

For comparison between results at the underdense regions and the average values in the universe, we chose 500 random points from all realisations and regarded the average values of these points as the reference values, i.e., the mean value of the universe.
For each radius $\theta_r$, we measured the average halo and peak counts along with the points.
Moreover, we obtained covariance matrices with the points for the calculations of $S/N$ defined in equation~(\ref{eq.sn}).
After employing more than 300 random points, the differences in the values of the covariance matrixes became a few percent.  

\section{RESULT}
Results for the halo number and weak lensing peak counts around the lowest peak are presented in this section, as well as calculations using the stacking method that determines the average properties around the underdense regions.

\label{sec.result}
\subsection{Halo mass function}
\label{sec.halo_pro}
We investigated the differences in the halo number count around large underdense regions as a function of halo mass by employing haloes whose masses were of the range $M=5\times10^{12}h^{-1}$M$_\odot$ to $10^{16}h^{-1}$M$_\odot$ and consequently dividing them into 15 bins in a log space.
Since we searched for the underdense regions from convergence maps whose source redshifts were 0.713, we performed the analysis for haloes whose redshift was less than 0.6. 

\begin{figure*}
\subfigure{\includegraphics[width=1.0\columnwidth]{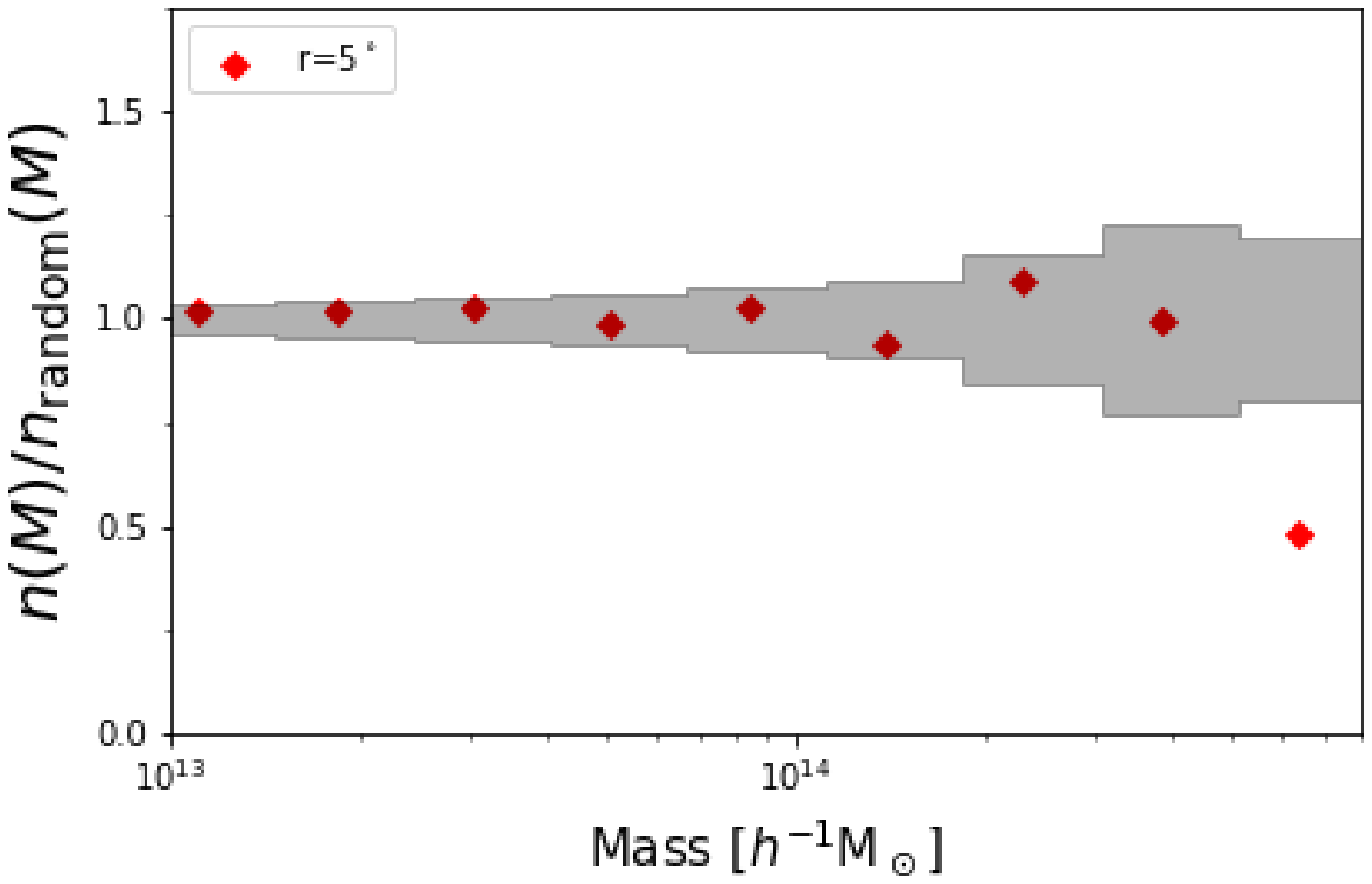}}
\subfigure{\includegraphics[width=1.0\columnwidth]{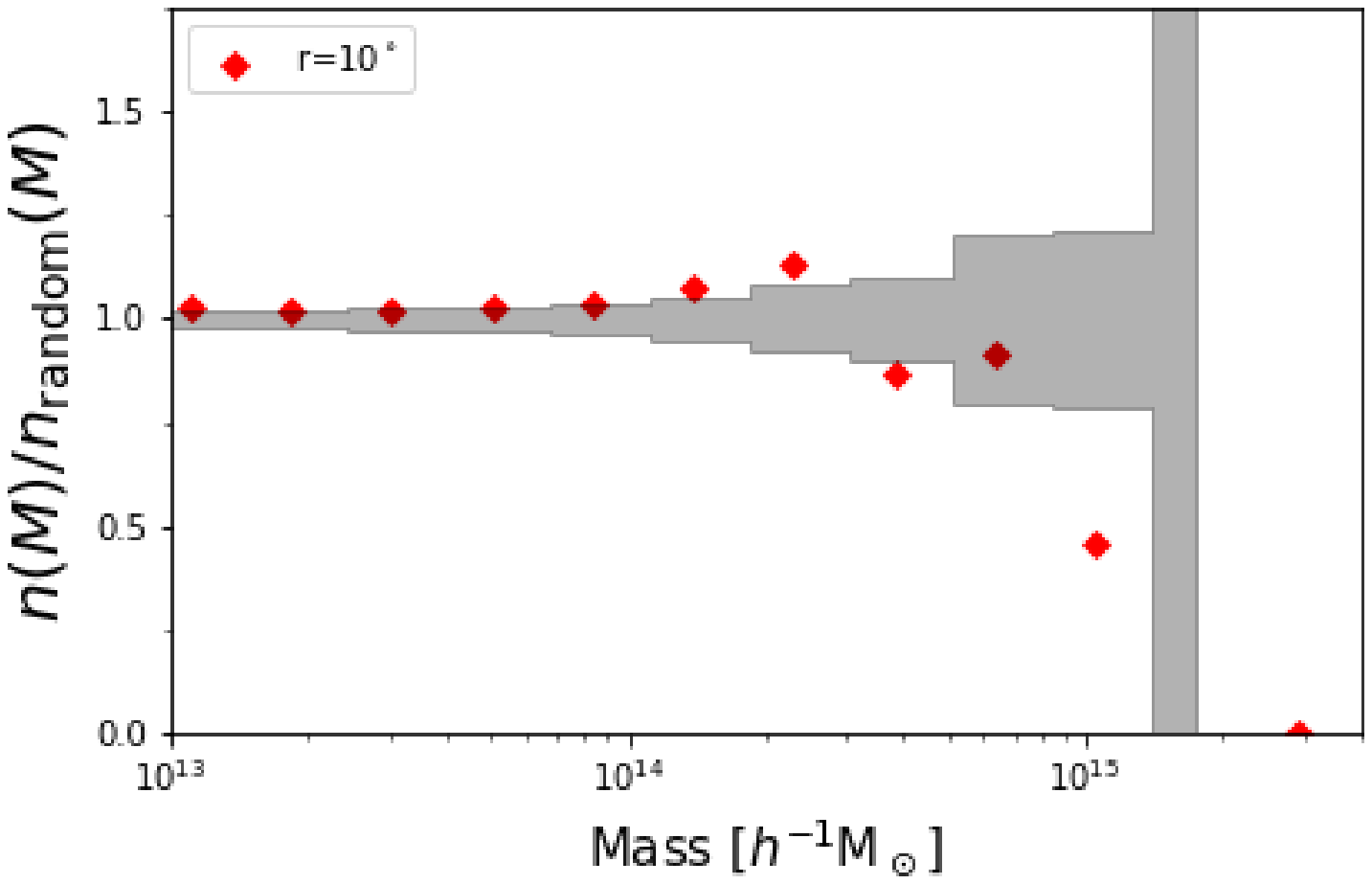}}
\subfigure{\includegraphics[width=1.0\columnwidth]{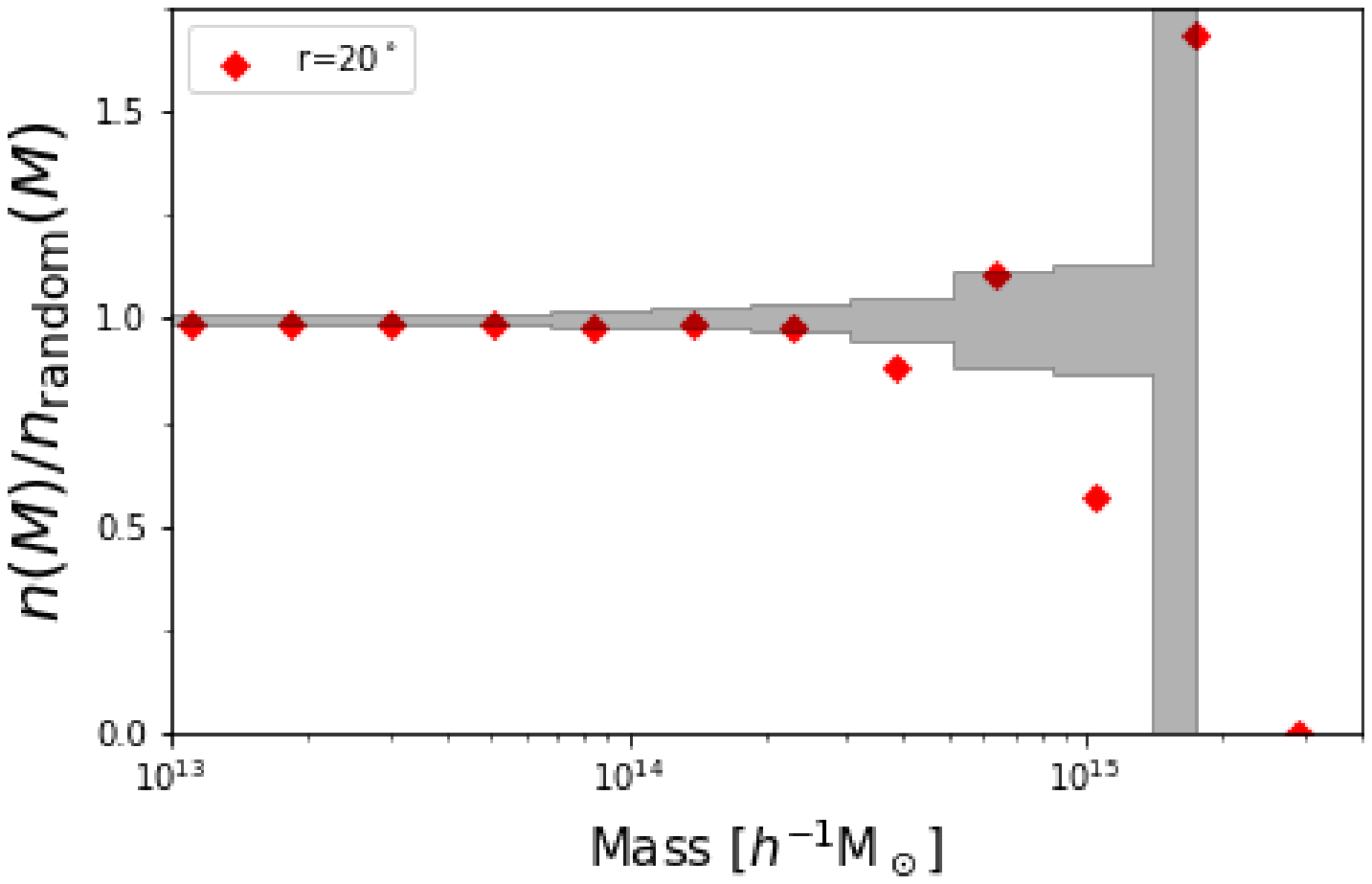}}
\subfigure{\includegraphics[width=1.0\columnwidth]{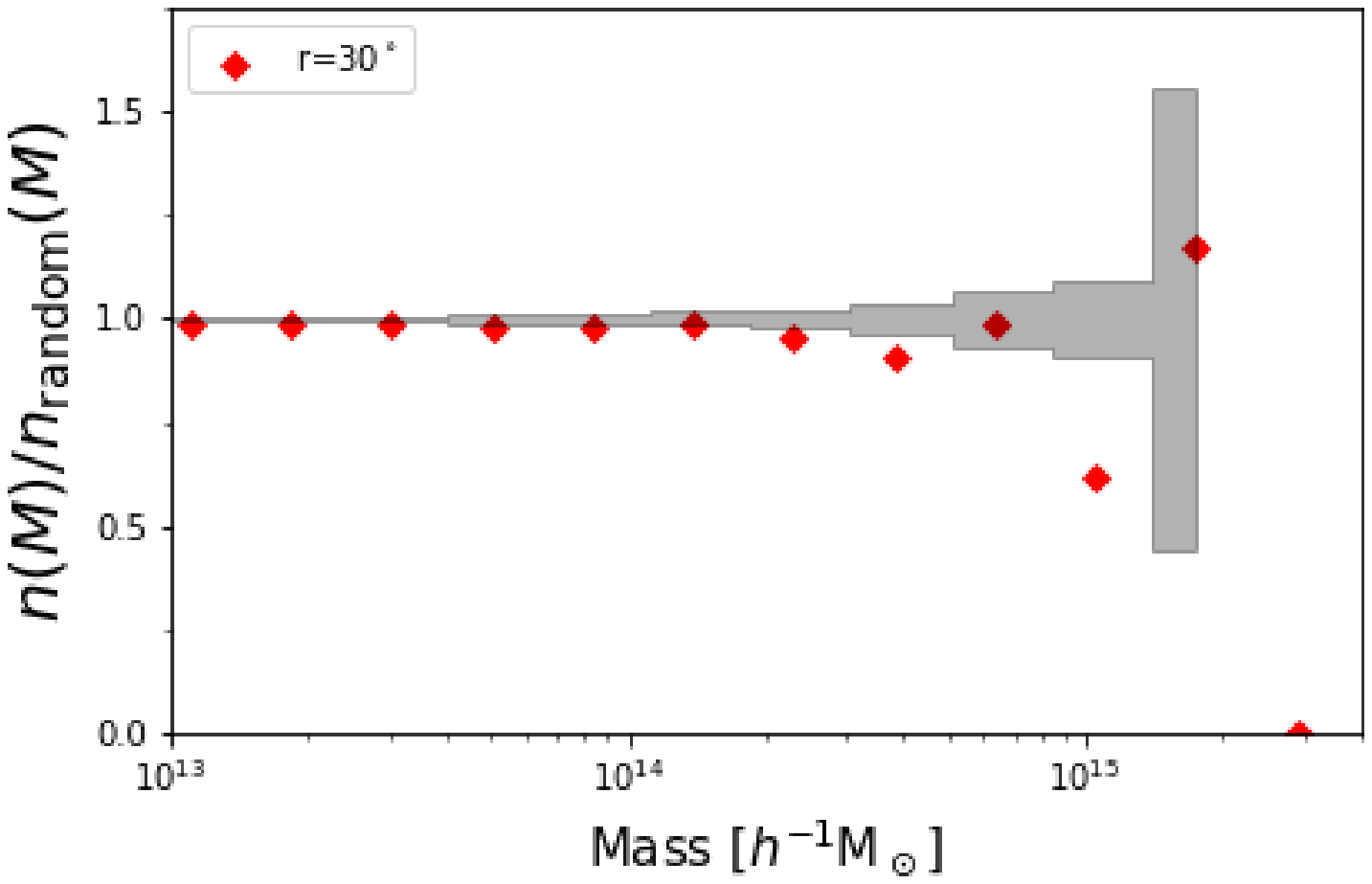}}
\subfigure{\includegraphics[width=1.0\columnwidth]{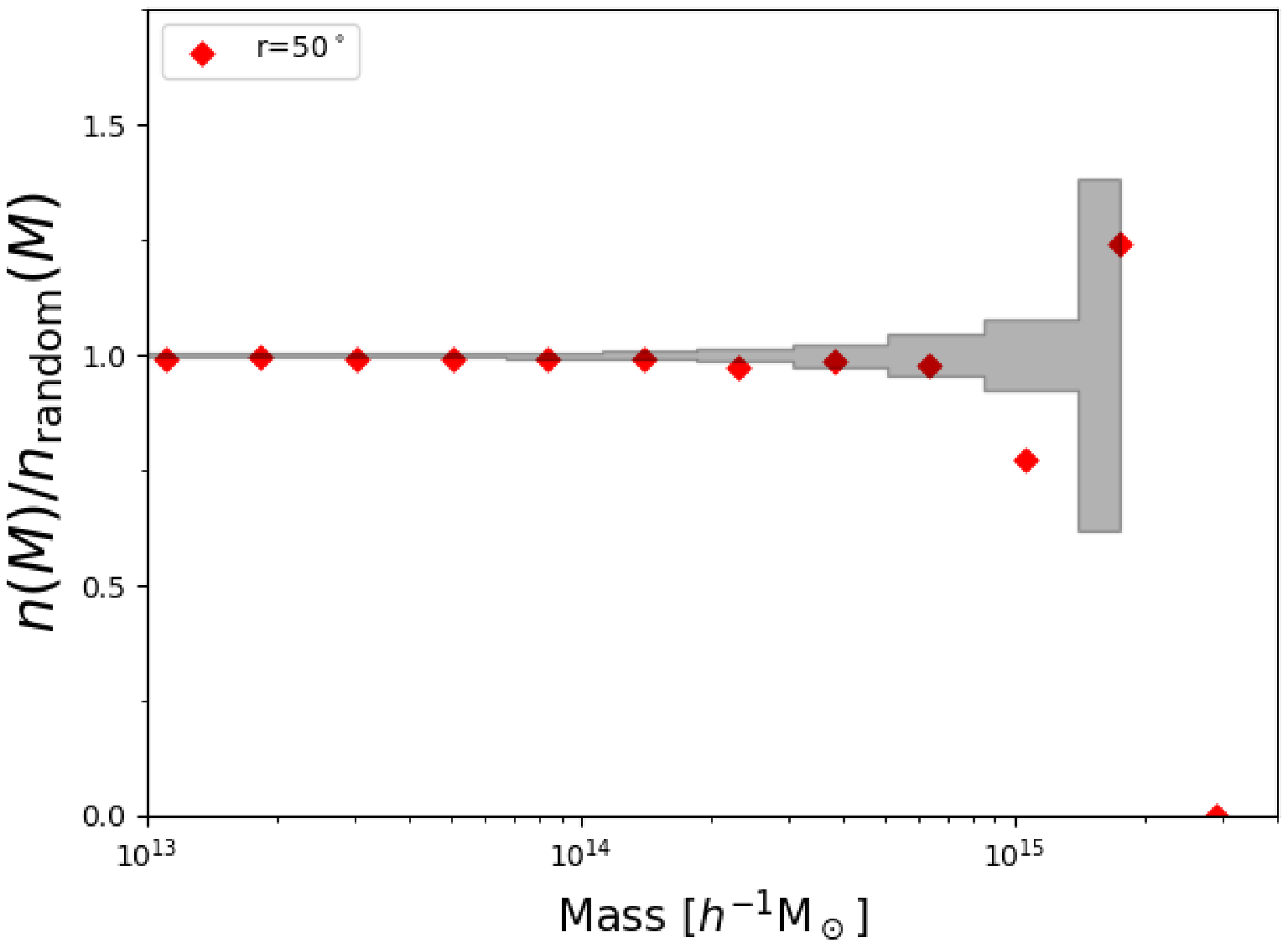}}
\caption{Differences in number count around the lowest convergence peak as a function of halo mass. 
The horizontal axis shows the halo mass, whereas the vertical axis depicts the ratio of the number of haloes at the peak to that at the random points.
Haloes at redshifts of $z\leq0.6$ are considered.
Red points represent measurements around the underdense region, and the shaded region indicates the standard deviation estimated from 500 random points.
Different panels show the measurement results for different search radii; {\it Top left:} $\theta_r=5$ degrees,
{\it Top right:} $\theta_r=10$ degrees, {\it Middle left:} $\theta_r=20$ degrees, {\it Middle right:} $\theta_r=30$ degrees 
and {\it Bottom:} $\theta_r=50$ degree.}
\label{fig.mass_peak}
\end{figure*}

\begin{figure*}
\subfigure{\includegraphics[width=1.0\columnwidth]{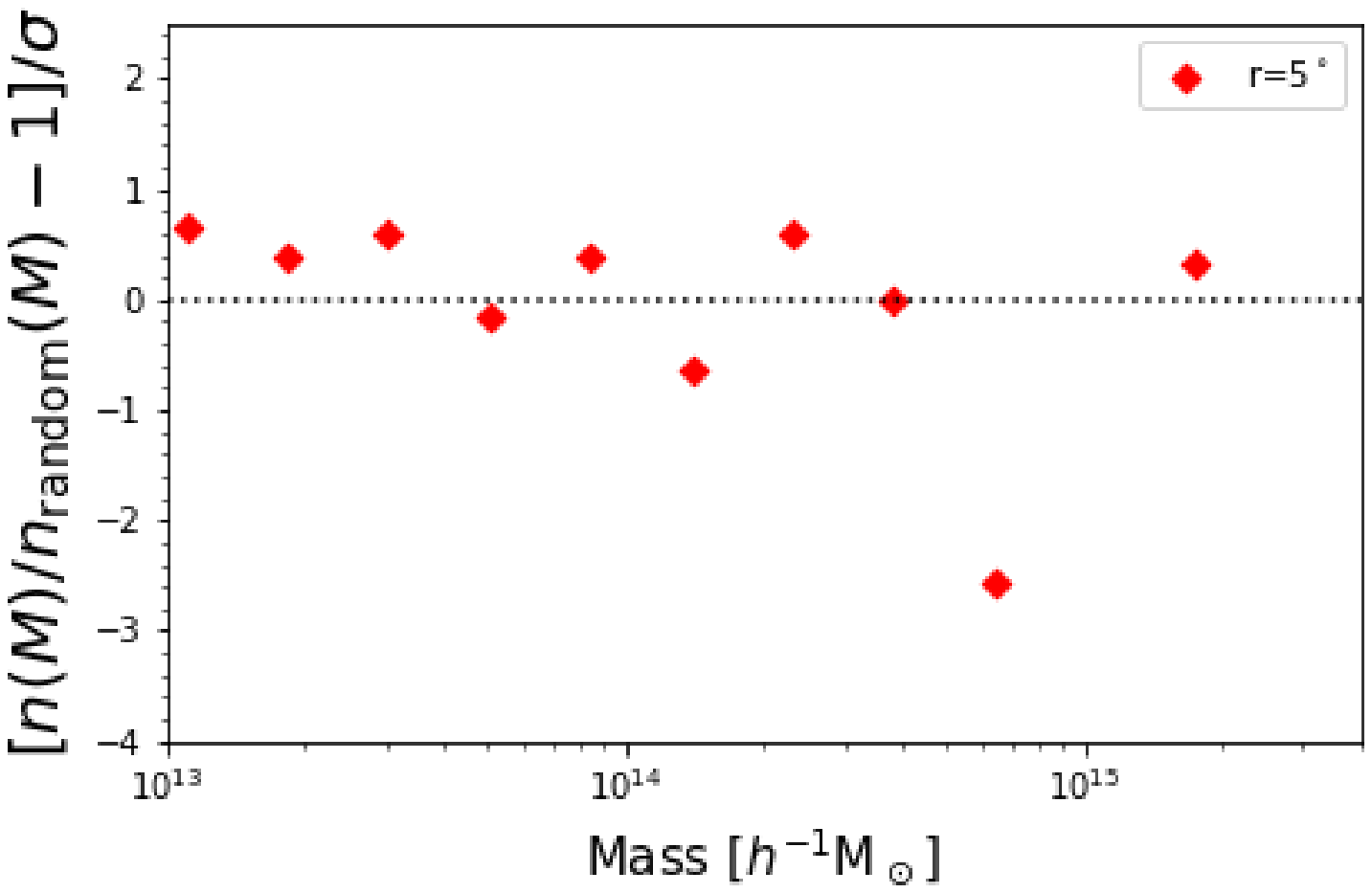}}
\subfigure{\includegraphics[width=1.0\columnwidth]{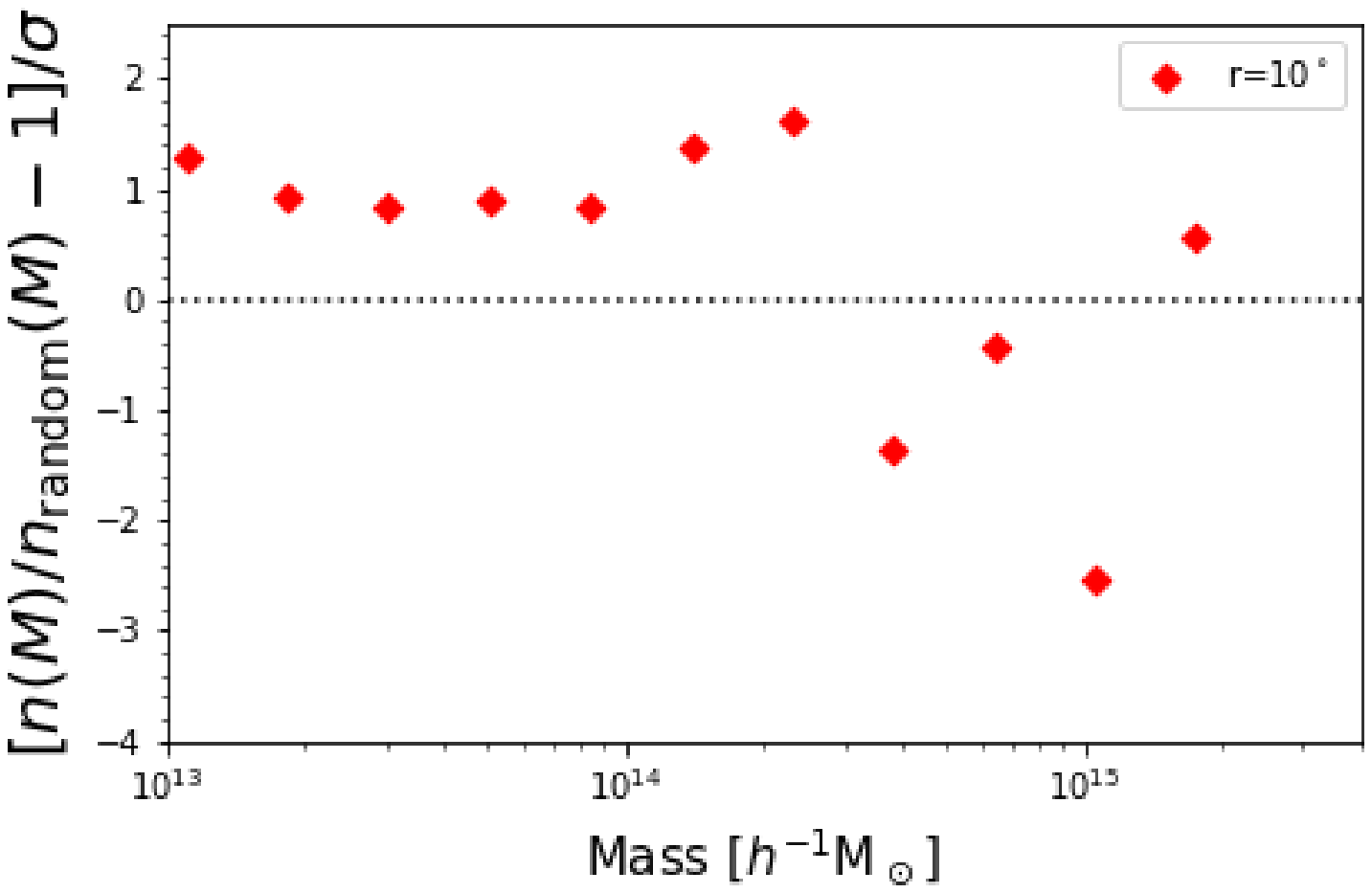}}
\subfigure{\includegraphics[width=1.0\columnwidth]{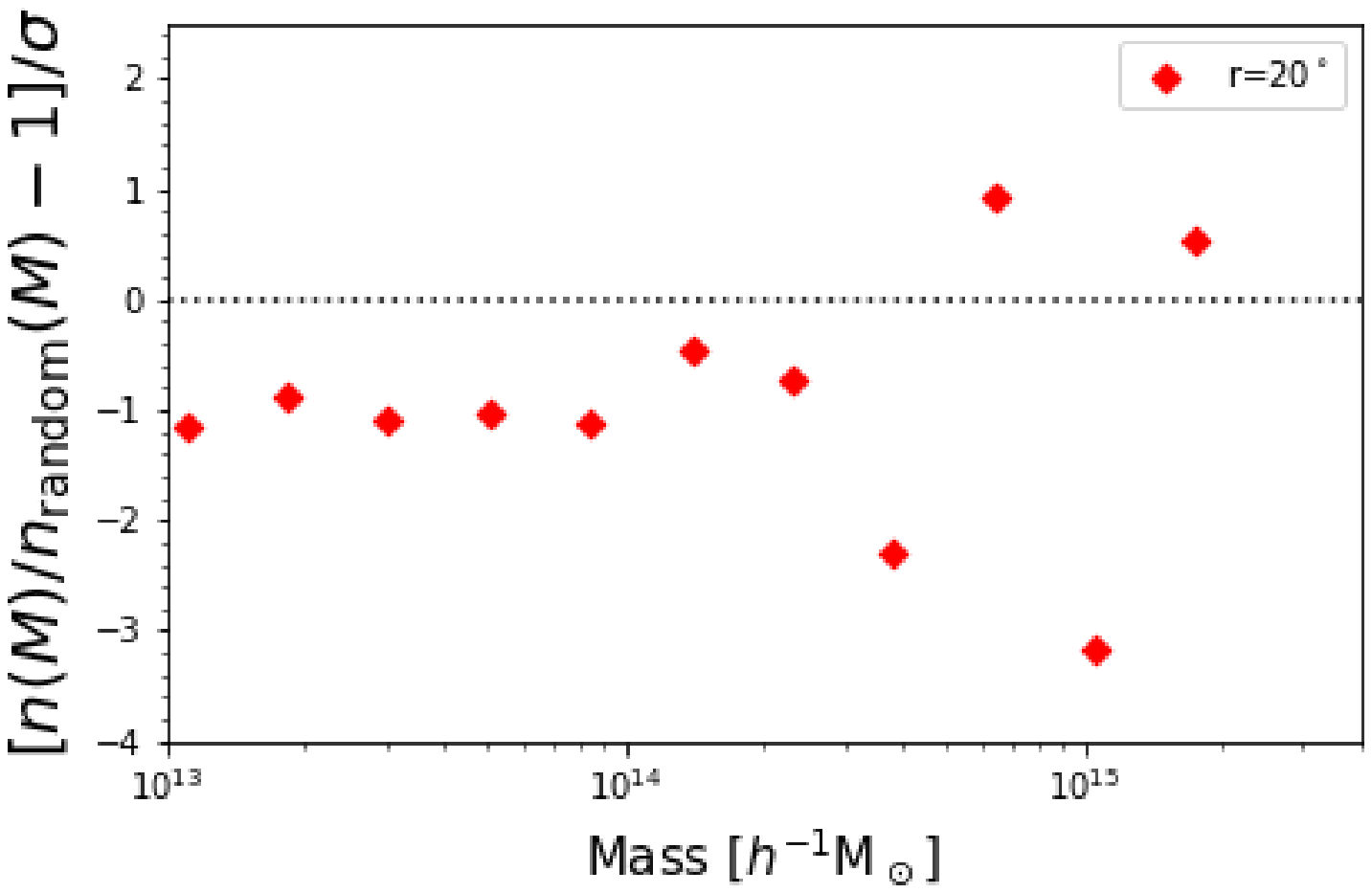}}
\subfigure{\includegraphics[width=1.0\columnwidth]{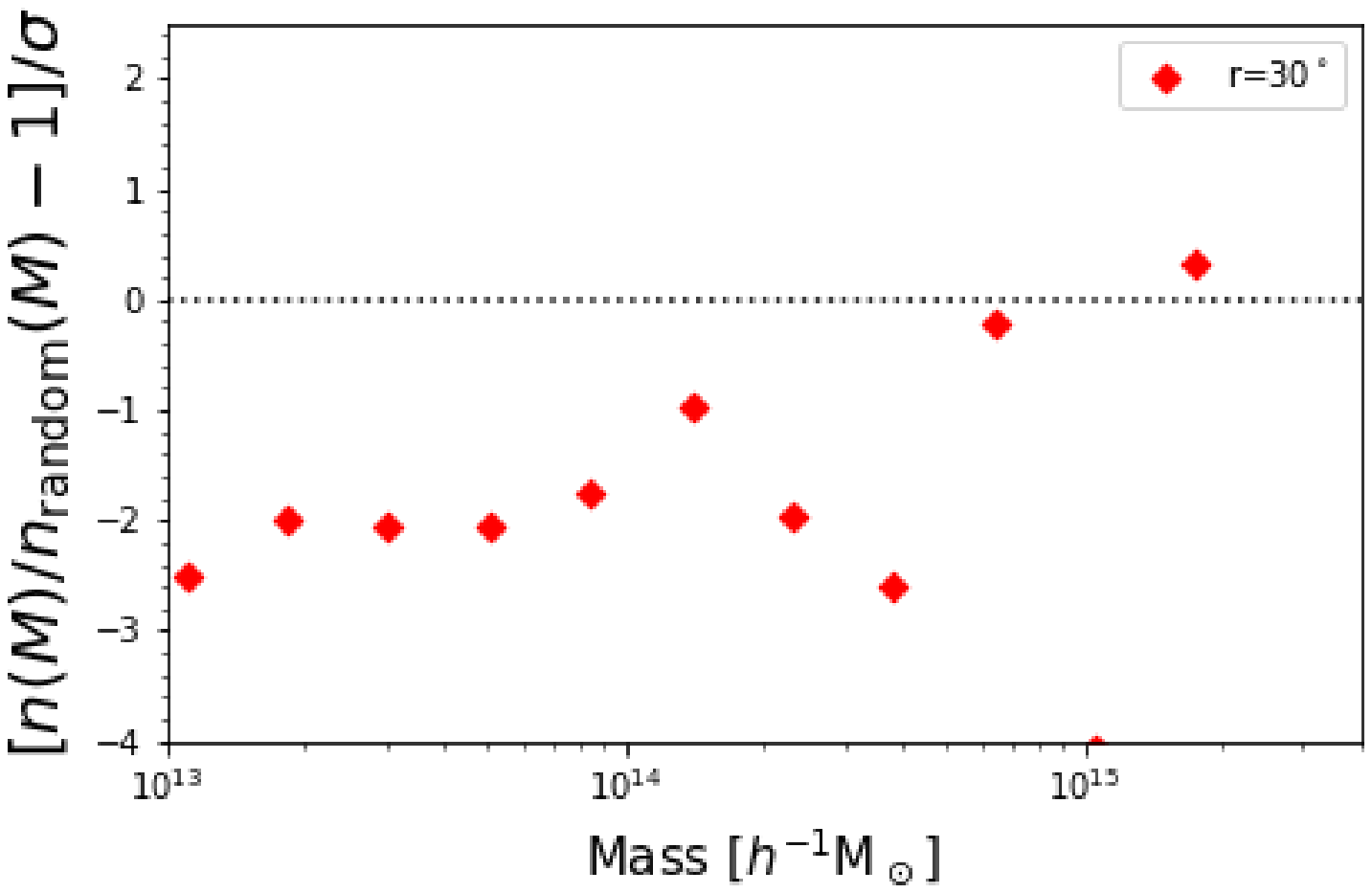}}
\subfigure{\includegraphics[width=1.0\columnwidth]{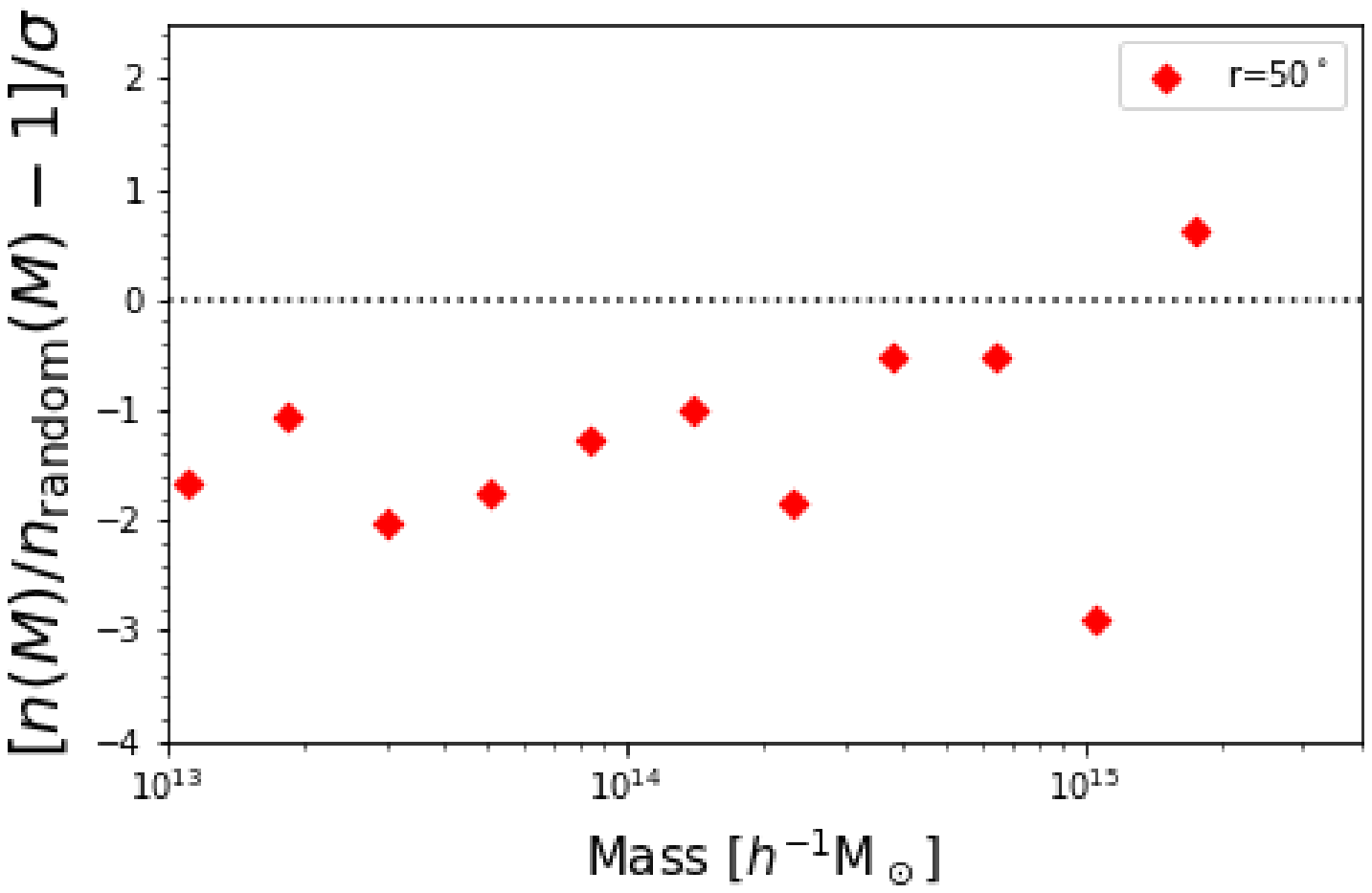}}
\caption{Significances of number count around the lowest convergence peak as a function of halo mass.
The horizontal axis shows a halo mass. 
The vertical axis illustrates the significance of the difference between the number count at the peak and the average value at the random points.
The standard deviation is calculated using the number counts at the random points in each bin. 
Haloes at redshifts of $z\leq0.6$ are considered in the analysis.
Each panel provides results for a given search radius $\theta_r$. 
{\it Top left:} $\theta_r=5$ degrees, {\it Top right:} $\theta_r=10$ degrees, {\it Middle left:} $\theta_r=20$ degrees, {\it Middle right:} $\theta_r=30$ degrees 
and {\it Bottom:} $\theta_r=50$ degrees.}
\label{fig.mass_peak_sn}
\end{figure*}

Figure~\ref{fig.mass_peak} shows the ratio of the number count at the lowest convergence peak to the average value at random points.
The red points indicated the number counts at the peak, whereas the shaded region depicted the $1\sigma$ standard deviation estimated from the 500 random points. 
Figure~\ref{fig.mass_peak_sn} shows the significance of the difference in the number count for each bin.
A significance $q$ at each bin is defined by 
\begin{equation}
q=\frac{ n(M)/n_{\rm random}(M) - 1}{\sigma},
\end{equation}
where $\sigma$ is the standard deviation at a bin, calculated with the random points.
Table~\ref{tab.sn_mass} shows the estimated total $S/N$ defined in equation~(\ref{eq.sn}) for several search radii centred at the lowest peak.
For the calculations, we excluded bins whose elements in the covariance matrix were equal to zero. 
As such, we can clearly see the effect of the decrement for the massive haloes wherein the significance of the differences is $\geq3$ in total. 

\begin{figure}
\subfigure{\includegraphics[width=1.0\columnwidth]{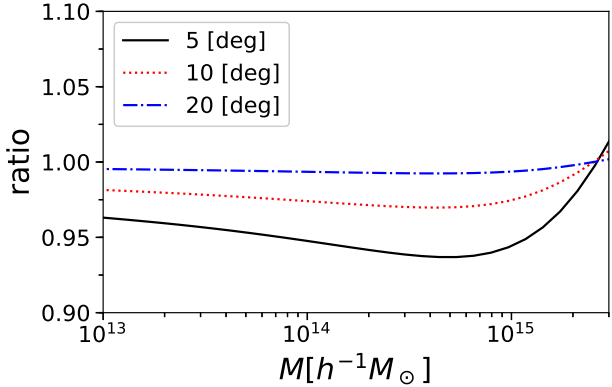}}
\caption{Differences in the number count of haloes at $z<0.6$ derived from the Press-Schechter theory. 
Each line corresponds to a number count within an aperture radius: {\it solid}, 10 degrees; {\it dotted}, 20 degrees; and {\it dotted-dashed}, 30 degrees.
The horizontal axis shows a halo mass. 
The vertical axis shows the ratio of the number of haloes toward two voids along the line of sight to that in fields.
The parameters (density contrast, comoving radius and redshift)  of the assumed voids are ($-0.16$, $63h^{-1}$Mpc, $0.23$) and ($-0.09$, $151h^{-1}$Mpc, 0.36). }
\label{fig.ps_mass_func}
\end{figure}

\begin{figure}
\subfigure{\includegraphics[width=1.0\columnwidth]{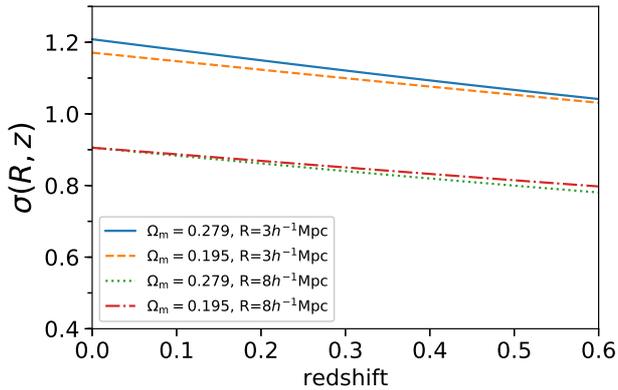}}
\caption{Differences in the variance of the density field for different cosmological models and scales as a function of redshift.
The vertical axis shows the rms variance of the linear density field on scale $R$.
The horizontal axis shows redshfit.
The solid, dashed, dotted and dash-dotted lines are the results for $(\Omega_{\rm m}, \textrm{R})=(0.279, 3h^{-1}{\rm Mpc})$, $(0.195, 3h^{-1}{\rm Mpc})$, $(0.279, 8h^{-1}{\rm Mpc})$ and $(0.195, 8h^{-1}{\rm Mpc})$, respectively.
In the calculation, we used the transfer function defined by equation~(\ref{eq.transfer})-(\ref{eq.betanode}).}
\label{fig.sigma_z}
\end{figure}

We compared the results around the largest underdense region to the predictions of the Press-Schechter theory by placing two top-hat voids along the line of sight with assumed properties measured by \citet{2018MNRAS.476..359H}.
The parameters (density contrast $\delta$, comoving radius $\chi$ and redshift $z$)  of the assumed voids were ($-0.16$, $63h^{-1}$Mpc, $0.23$)  and ($-0.09$, $151h^{-1}$Mpc, $0.36$).
We assumed that the density contrast of the voids was constant inside the void radii, i.e., top-hat voids.
We estimated the number of haloes inside the voids by modifying the matter density parameter defined in equation~(\ref{eq.omega_m}) and ignored the effect from the ridge regions around the voids.
Under these conditions, the volume of the voids occupies a few percent of the total volume in the line of sight down to redshift $z\leq0.6$ within a search radius. 
On the other hand, from equation~(\ref{eq.massfunc}) the number of haloes inside the voids decreases $\sim20$ percent.
Figure~\ref{fig.ps_mass_func} shows the differences in the number count of haloes estimated via equation~(\ref{eq.massfunc}) as a function of halo mass;
the solid, dotted and dotted-dashed lines represent the results for different search radii of 10, 20 and 30 degrees, respectively.
The vertical axis shows the difference between the number count of haloes estimated with equation~(\ref{eq.omega_m}) by assuming the two voids in the line of sight and that without the voids.
Moreover, the Press-Schechter theory predicts a few percent decrement in the number count toward the measured underdense regions, which is broadly consistent with our result for the low mass haloes in the simulation. 
For massive haloes, the number count in the voids predicted from the Press-Schechter mass function increases.
Such a change reflects the variation in the matter power spectrum.
Figure~\ref{fig.sigma_z} shows the rms variance of the density field for different cosmological models and scales.
When the matter density decreases, the power spectrum at large wavenumber is suppressed due to the suppression of the transfer function at the present time. 
This effect suppresses the growth of matter on small scales.
On the other hand, the value of the growth factor at a certain time is larger than those in higher density universes, because the change in the growth factor is smaller in lower matter density universes.
Since the latter effect is stronger than the former effect, the number count in that scale increases. 

\begin{table}
\caption{Singal-to-noise ratios for halo number counts at the lowest peak for each search radius $\theta_r$. 
Column~(1): search radius. column~(2): total $S/N$ estimated via equation~(\ref{eq.sn}): and column~(3): number of bins.}
\begin{center}
\begin{tabular}{|c|c|c|}
radius $\theta_r$ [degree] & S/N  & number of bins\\ \hline\hline
5                                        & 2.52 & 13\\	
10       				& 2.75 & 13\\
20					& 2.57 & 13\\
30					& 3.39& 13\\
50					& 3.26 & 13
\end{tabular}
\end{center}
\label{tab.sn_mass}
\end{table}

\begin{figure*}
\subfigure{\includegraphics[width=1.0\columnwidth]{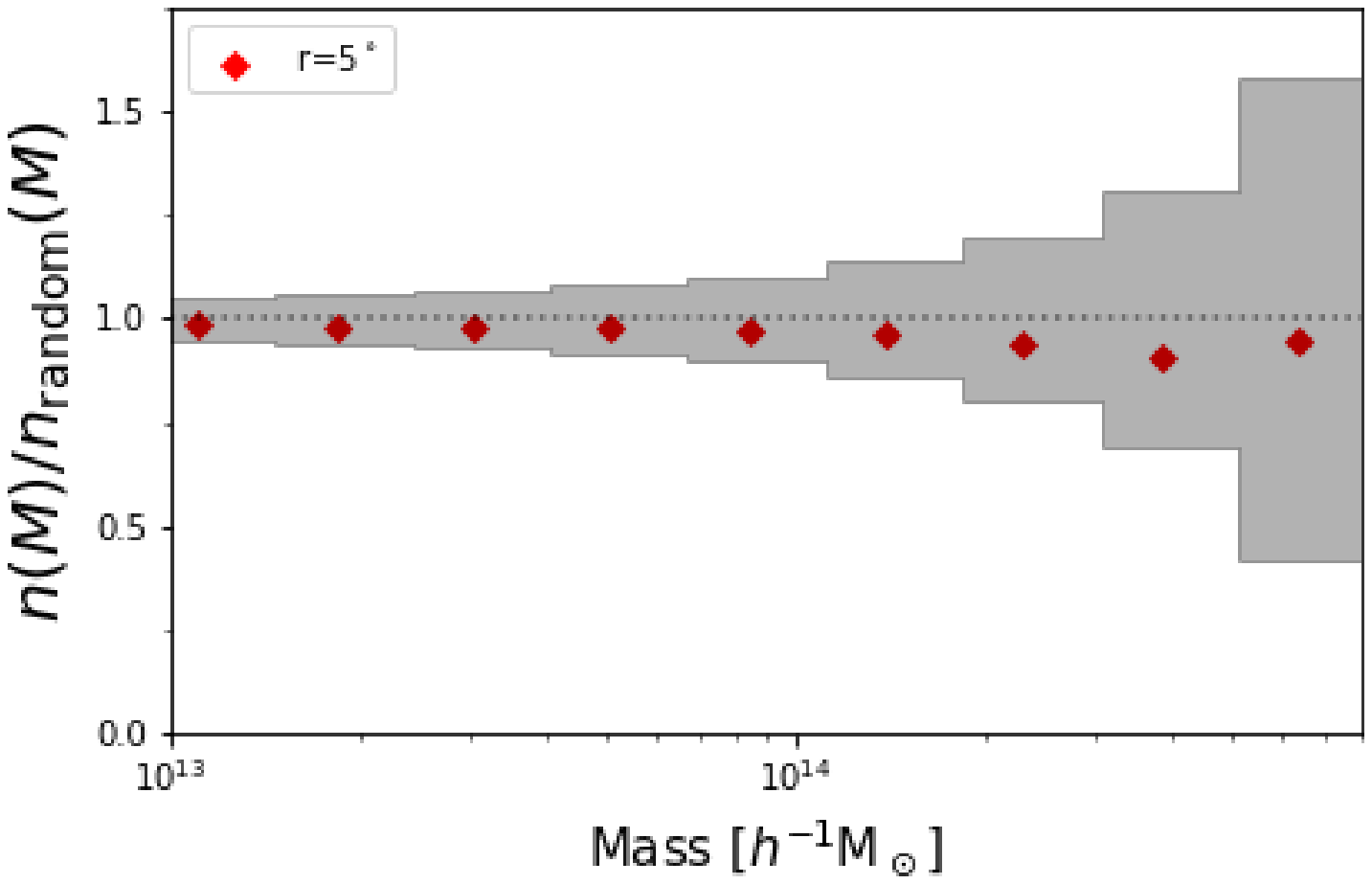}}
\subfigure{\includegraphics[width=1.0\columnwidth]{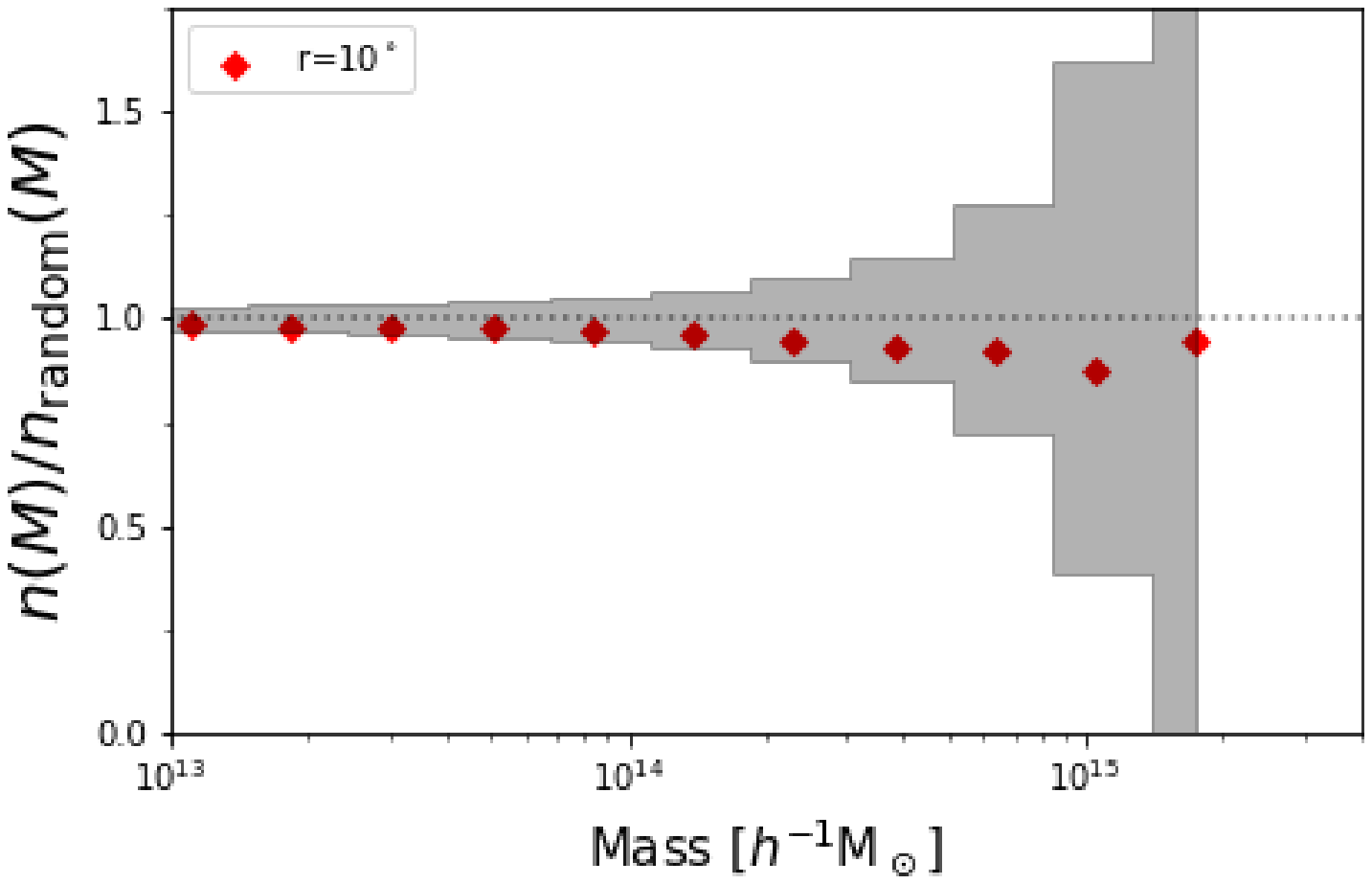}}
\subfigure{\includegraphics[width=1.0\columnwidth]{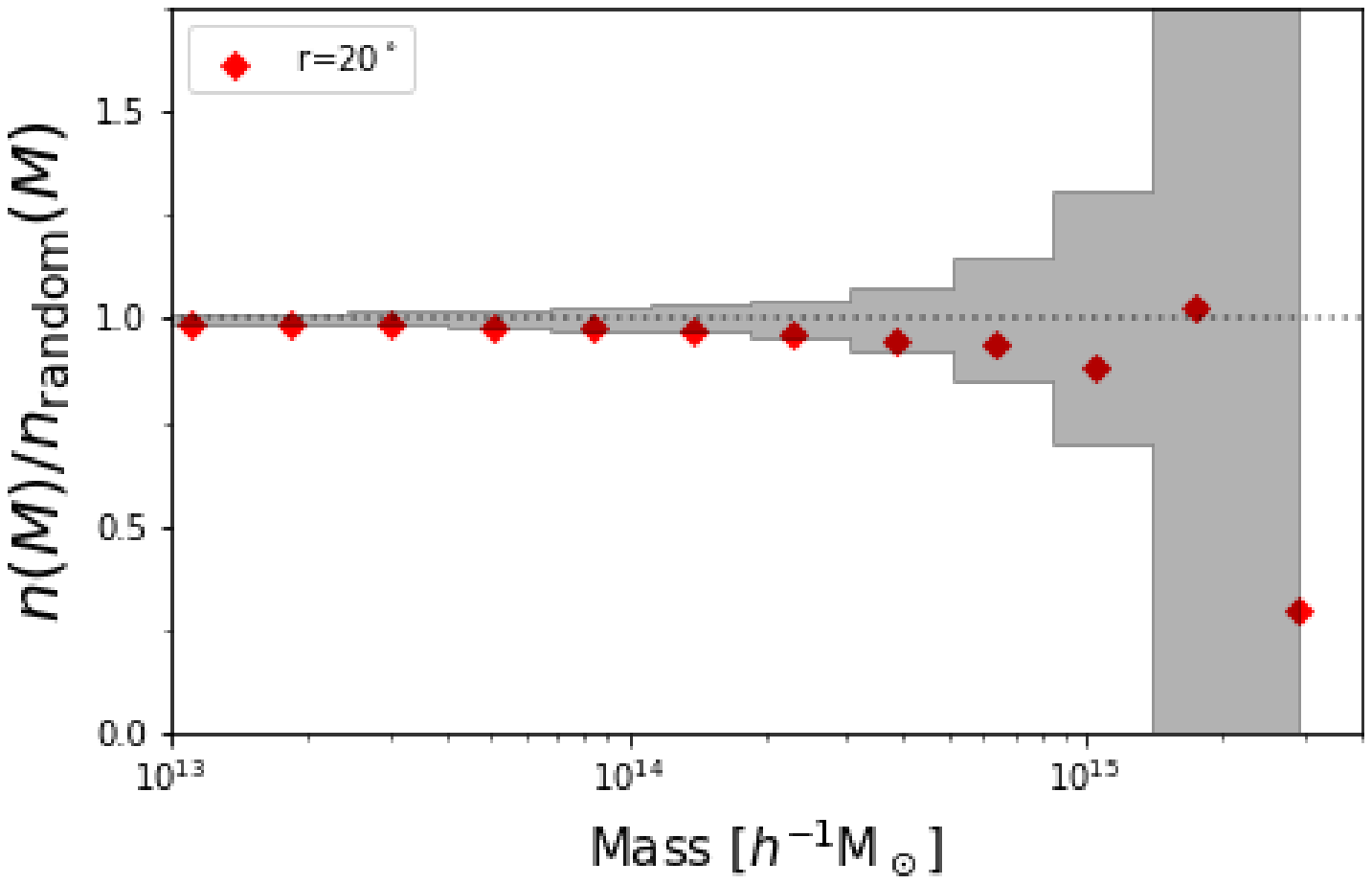}}
\subfigure{\includegraphics[width=1.0\columnwidth]{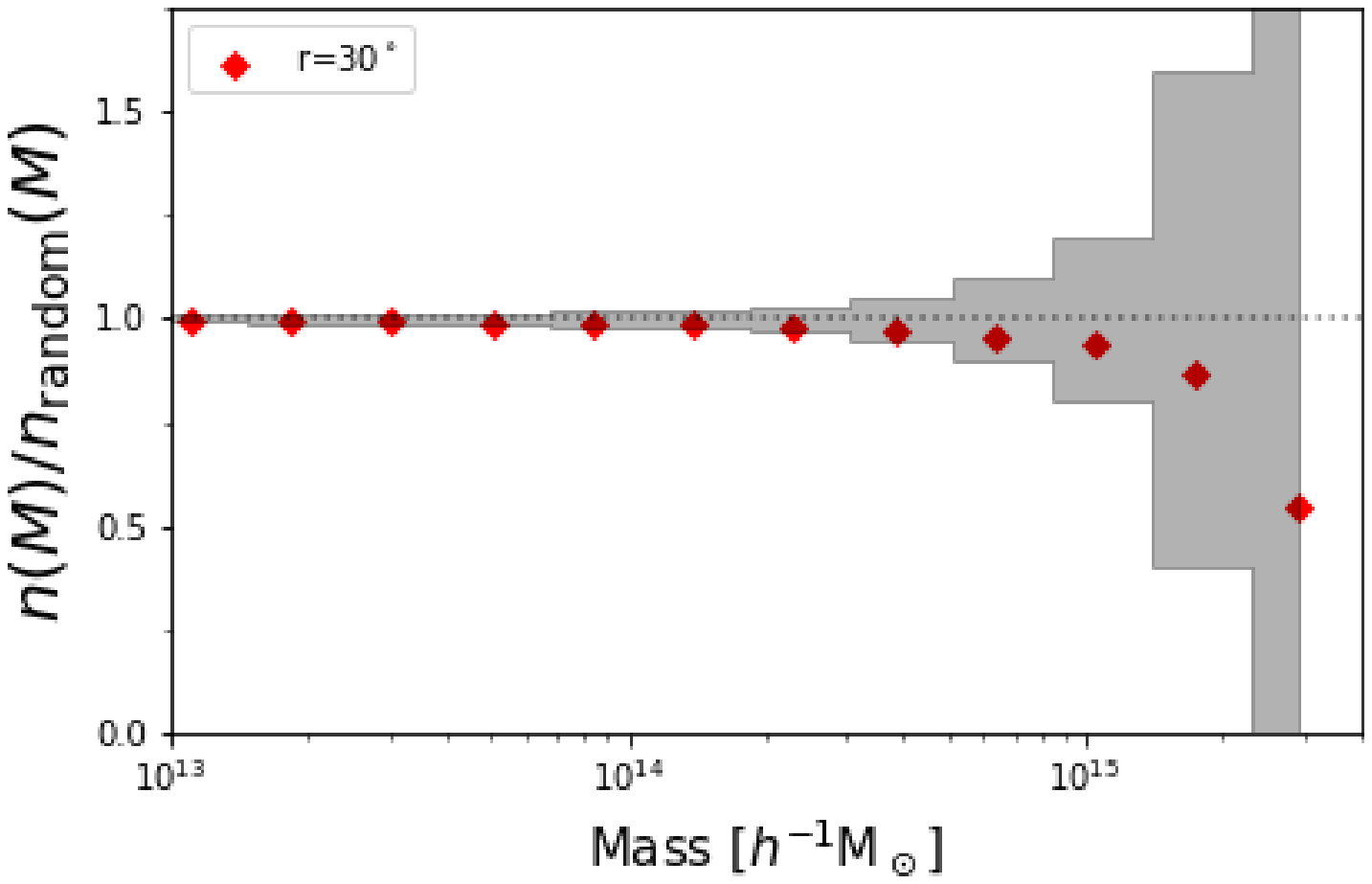}}
\subfigure{\includegraphics[width=1.0\columnwidth]{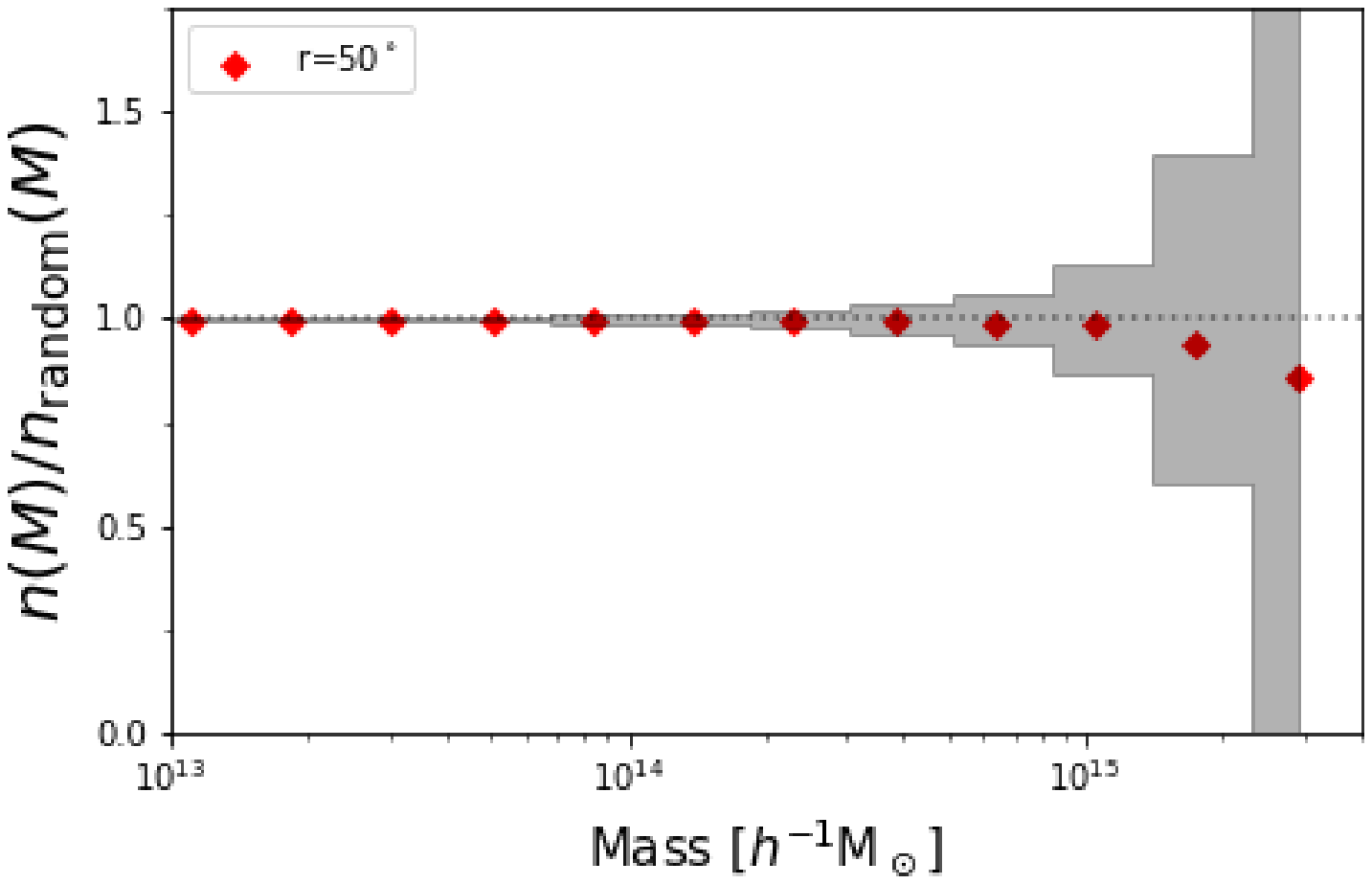}}
\caption{Similar to figure~\ref{fig.mass_peak}, but results are obtained by stacking the low density regions.}
\label{fig.mass_stack}
\end{figure*}

\begin{figure*}
\subfigure{\includegraphics[width=1.0\columnwidth]{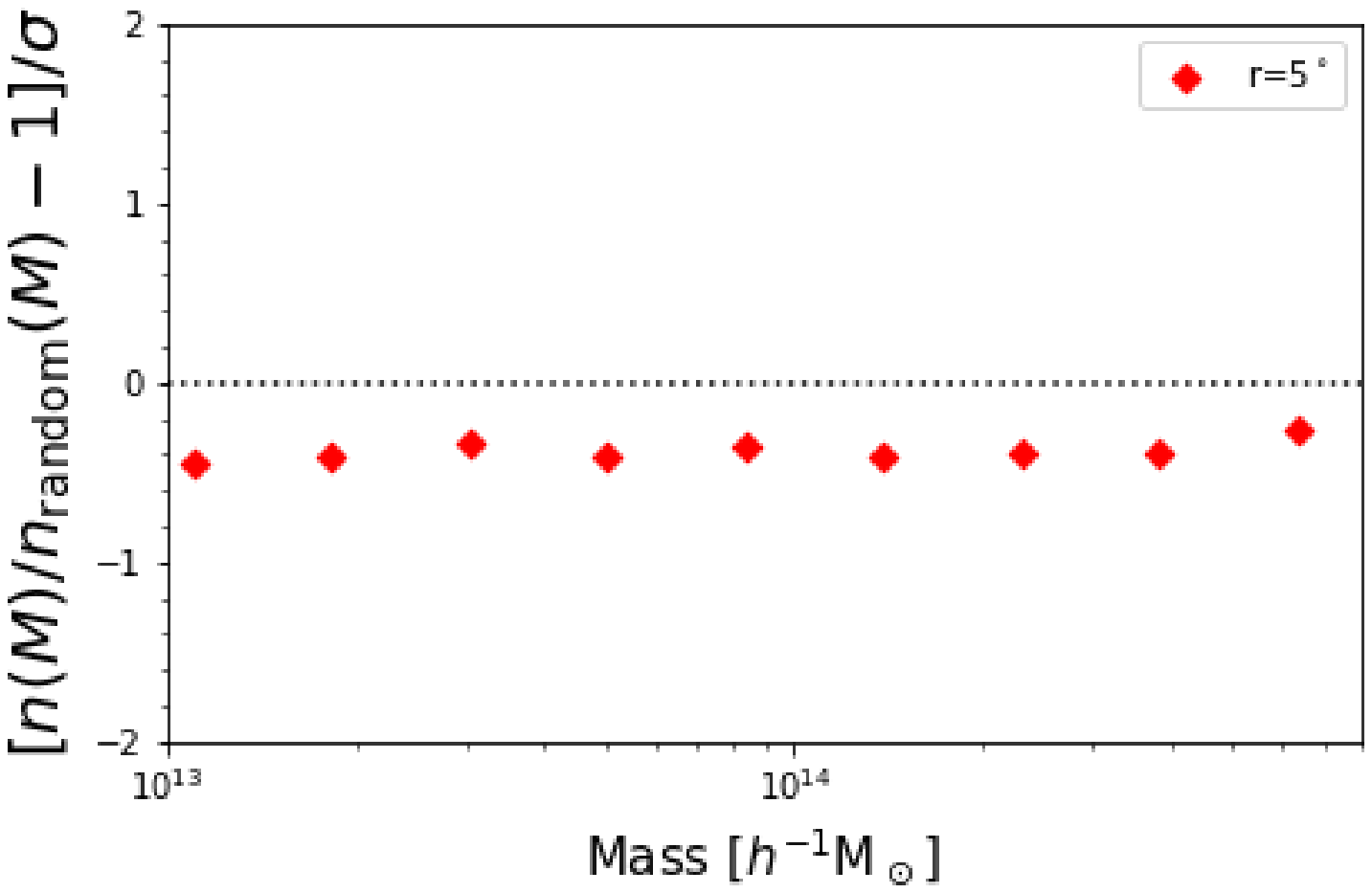}}
\subfigure{\includegraphics[width=1.0\columnwidth]{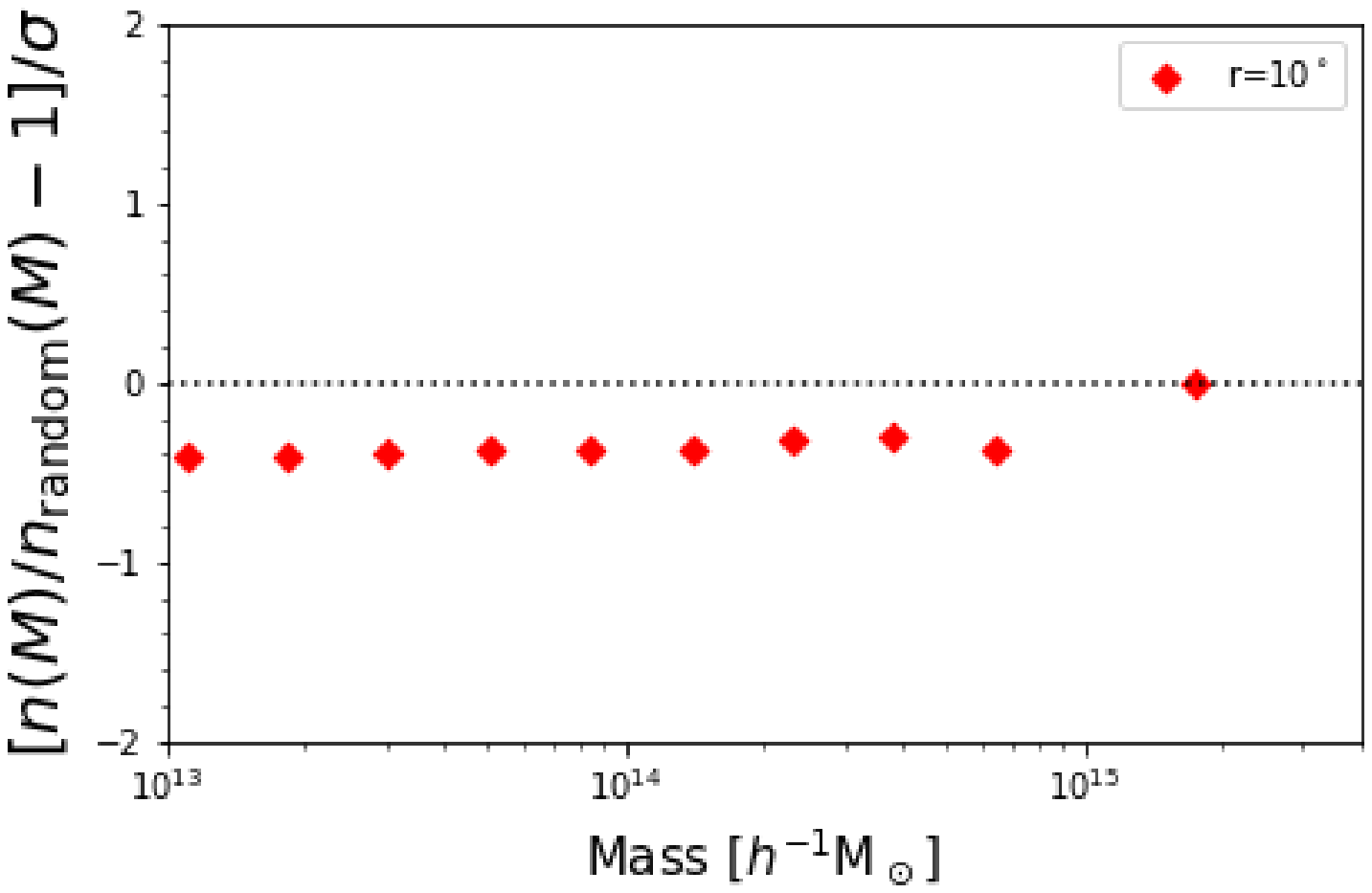}}
\subfigure{\includegraphics[width=1.0\columnwidth]{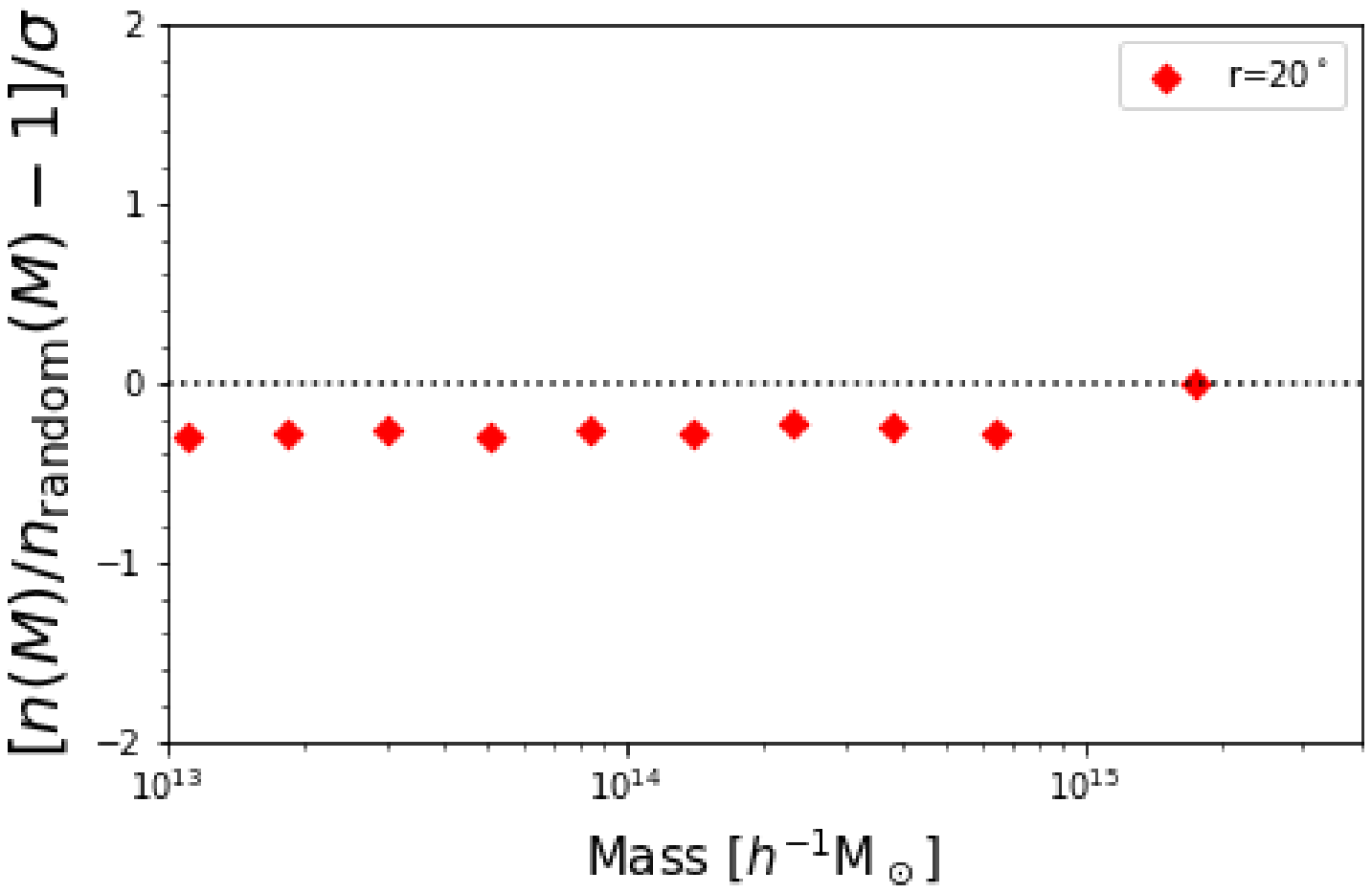}}
\subfigure{\includegraphics[width=1.0\columnwidth]{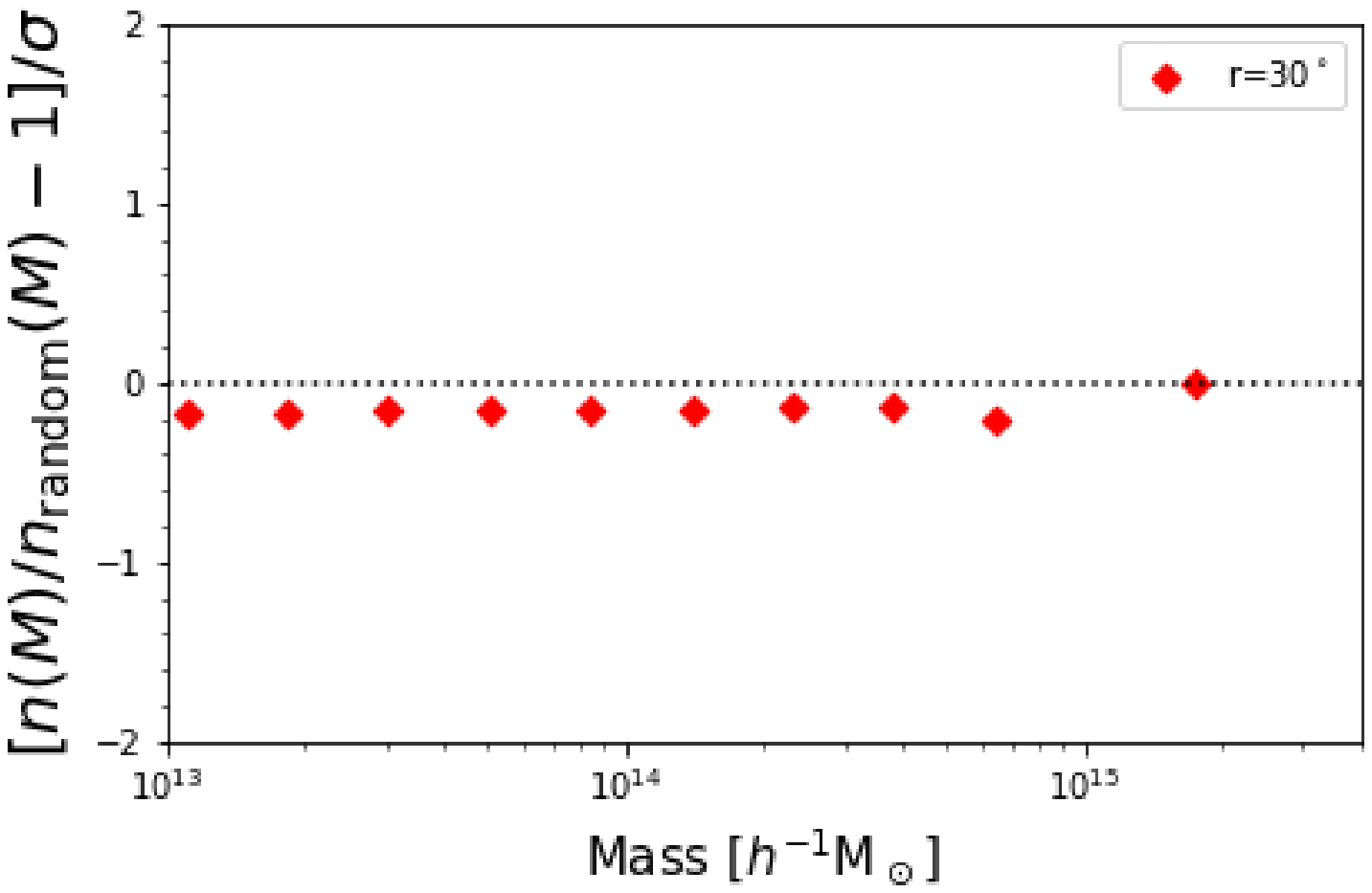}}
\subfigure{\includegraphics[width=1.0\columnwidth]{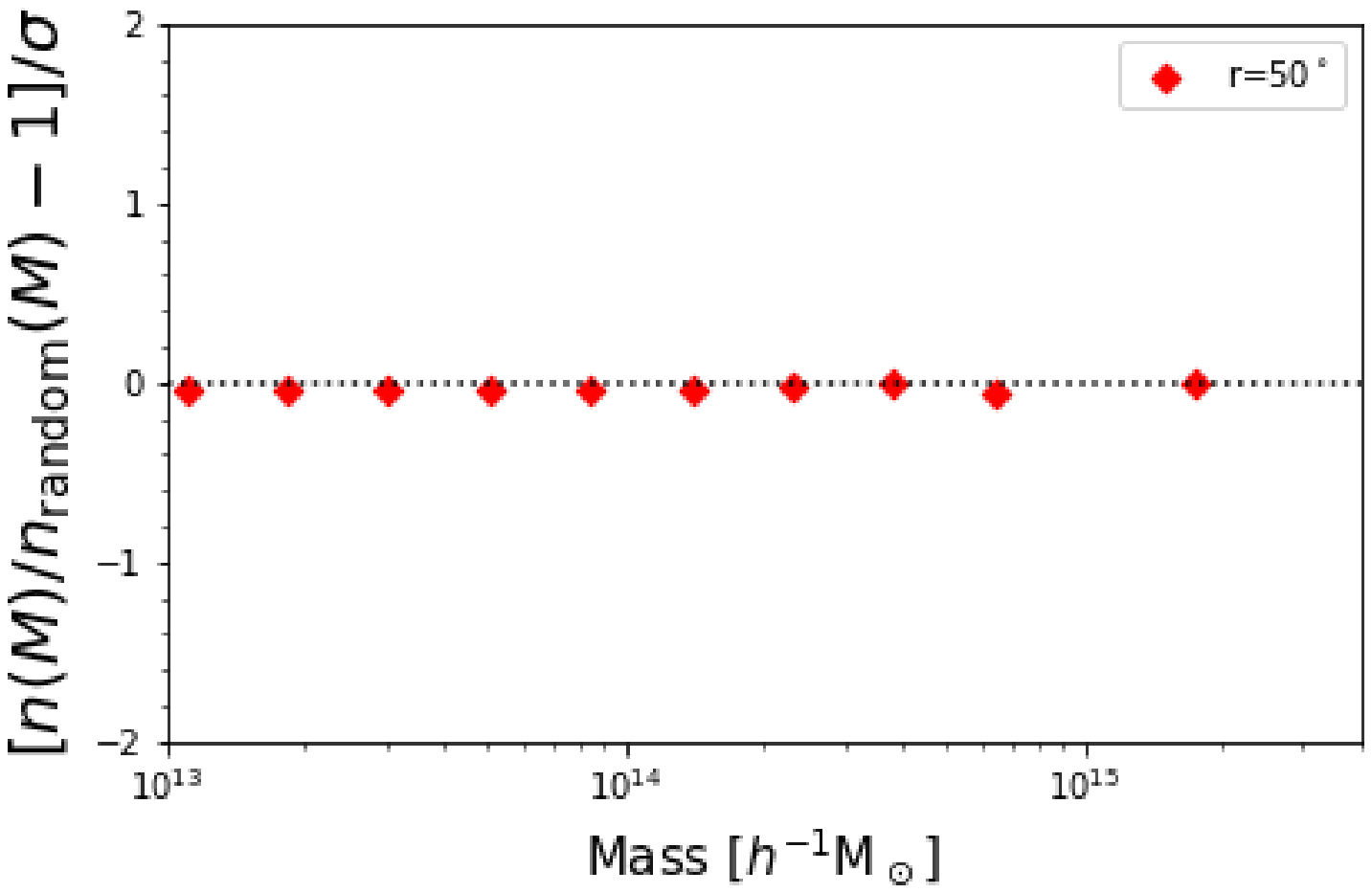}}
\caption{Similar to figure~\ref{fig.mass_peak_sn}, but results are for the stacked analysis.}
\label{fig.stack_sn}
\end{figure*}

Figures~\ref{fig.mass_stack} and \ref{fig.stack_sn} show the halo number count and the significance of the number count calculated with a stacking method around the underdense regions. 
The red points indicate the average values estimated with the 106 underdense regions whose lowest peak values are lower than -10 as defined in equation~(\ref{eq.sn_peak}).
$S/N$ ratios for the stacked results are shown in Table~\ref{tab.sn_mass_stack}.
Although the environmental effect of the decrement is not conspicuous, a similar behavior is observed in the lowest peak, indicating that growth of haloes is suppressed in the large underdense regions, 
a feature consistent with previous simulation results describing that massive haloes reside in denser regions \citep{1999MNRAS.304..767S, 2002MNRAS.329..61S, 2006MNRAS.367.1039H, 2009MNRAS.398.1742H, 2010ApJ...712..484F, 2018arXiv180604684A}.

\begin{table}
\caption{Similar to table~\ref{tab.sn_mass}, but the results are for the stacking analysis.}
\begin{center}
\begin{tabular}{|c|c|c|}
radius $\theta_r$ [degree] & S/N  & number of bins\\ \hline\hline
5                                        & 0.49 & 13\\	
10       				& 0.92 & 13\\
20					& 1.38 & 13\\
30					& 1.20 & 13\\
50					& 0.50 & 13
\end{tabular}
\end{center}
\label{tab.sn_mass_stack}
\end{table}

\subsection{Peak statistics}
Furthermore, we investigated the environmental effects from large underdense regions on weak lensing peak statistics in the entire simulation through peak count measurements around the lowest peak and average value calculations around the 106 negative peaks. 
Afterwords, we divided $S/N$ from -20 to 20 into 80 bins, i.e., $\Delta\nu=0.5$.

\begin{figure*}
\subfigure{\includegraphics[width=1.0\columnwidth]{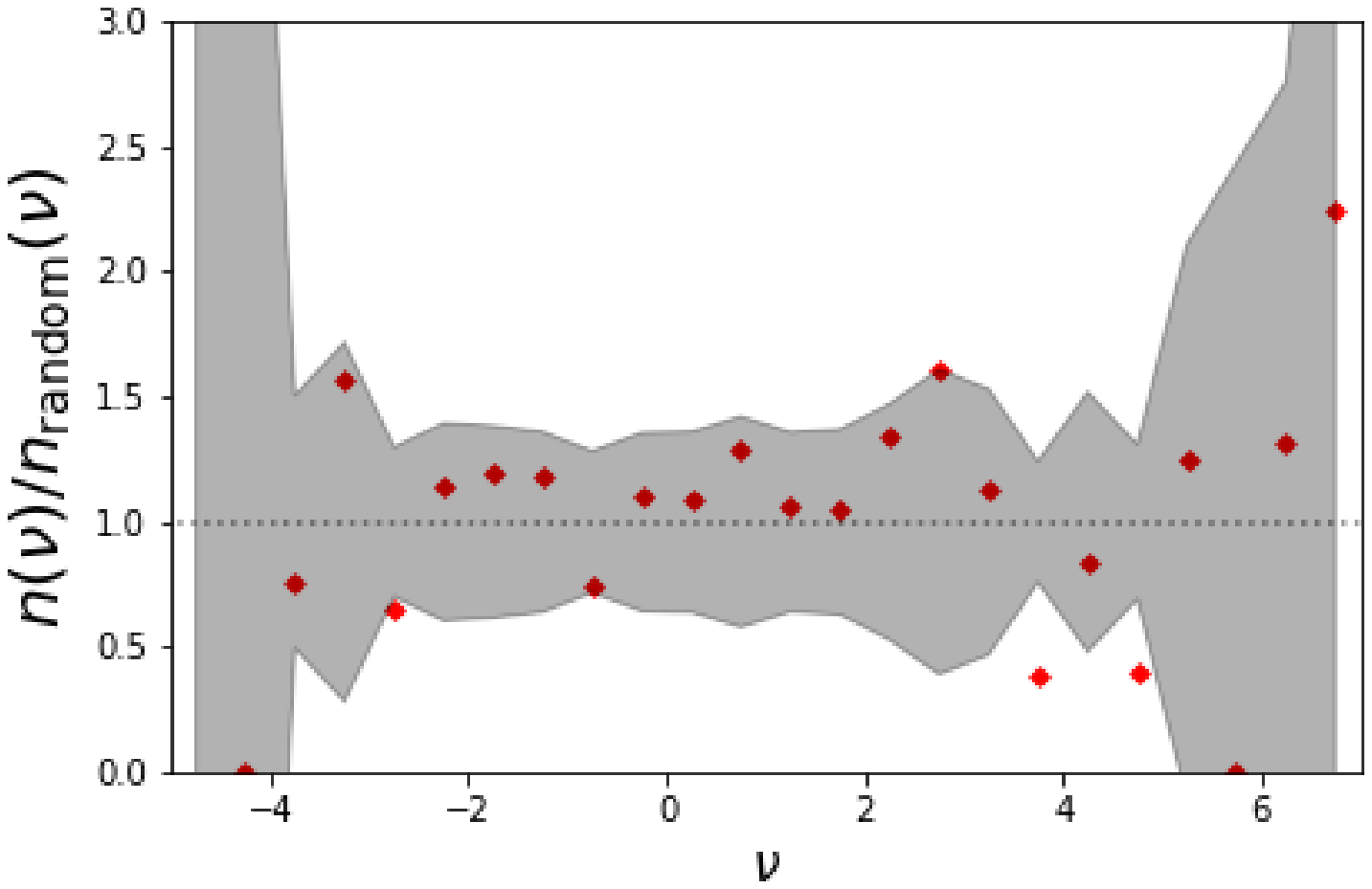}}
\subfigure{\includegraphics[width=1.0\columnwidth]{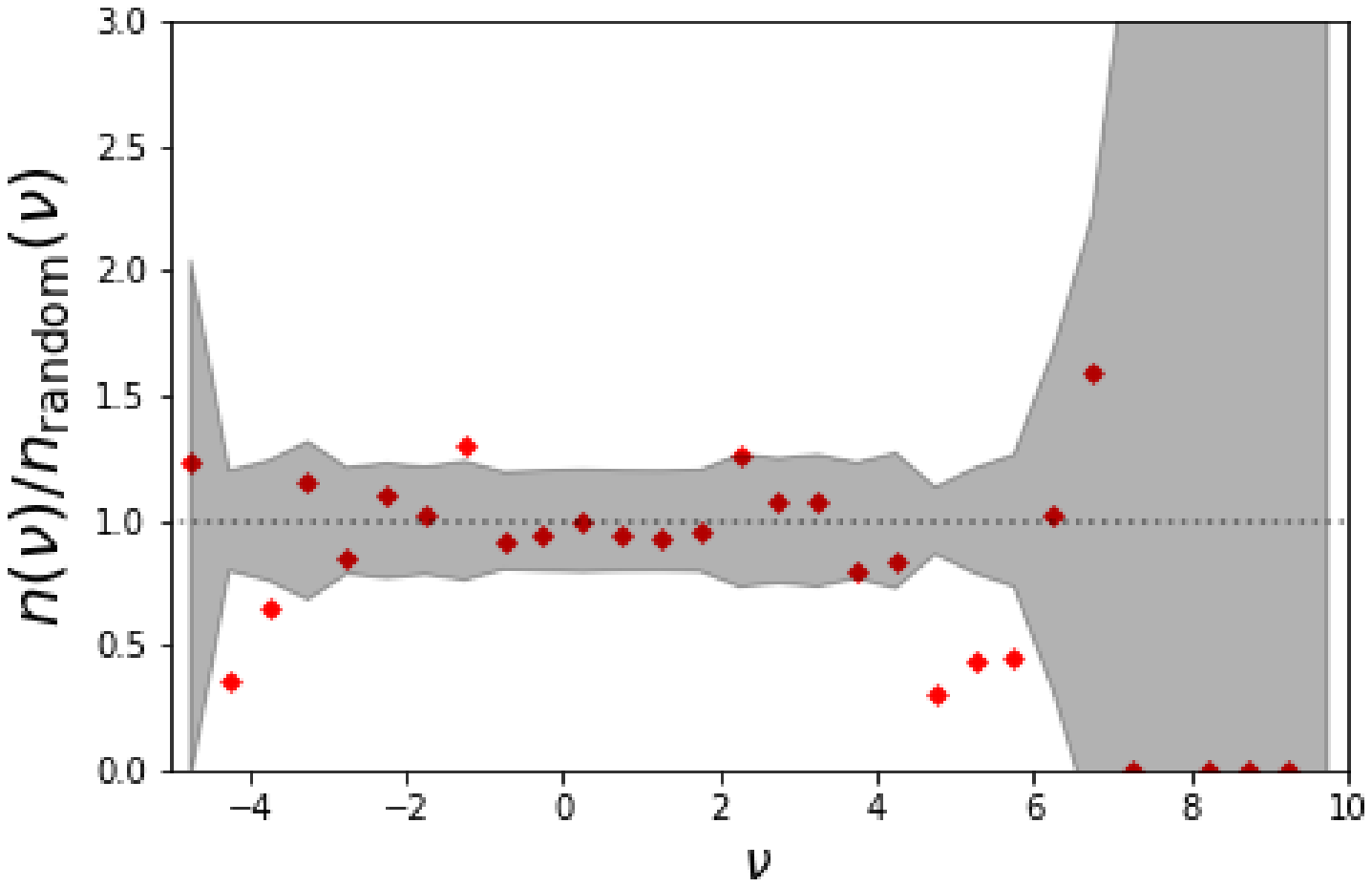}}
\subfigure{\includegraphics[width=1.0\columnwidth]{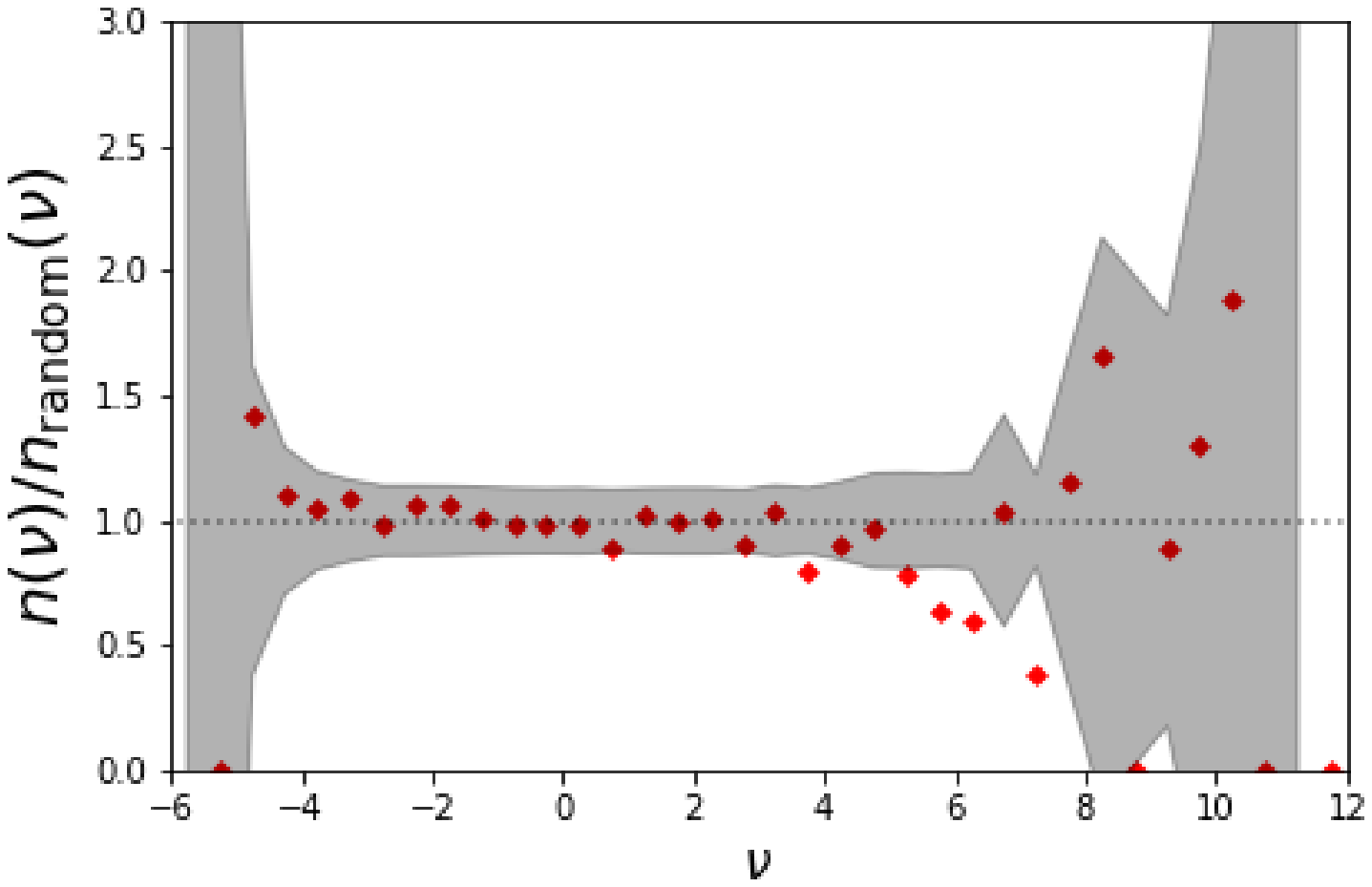}}
\subfigure{\includegraphics[width=1.0\columnwidth]{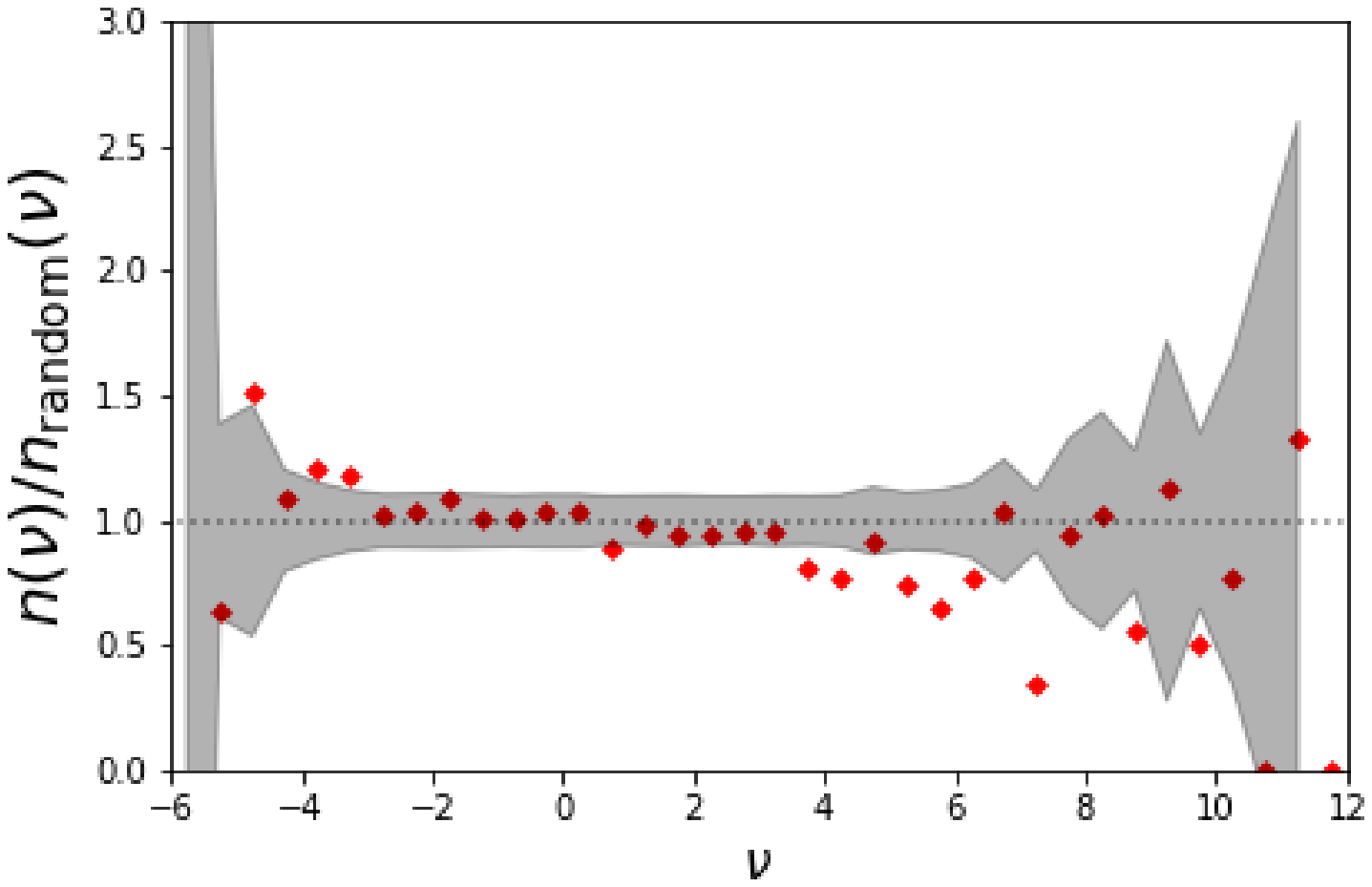}}
\subfigure{\includegraphics[width=1.0\columnwidth]{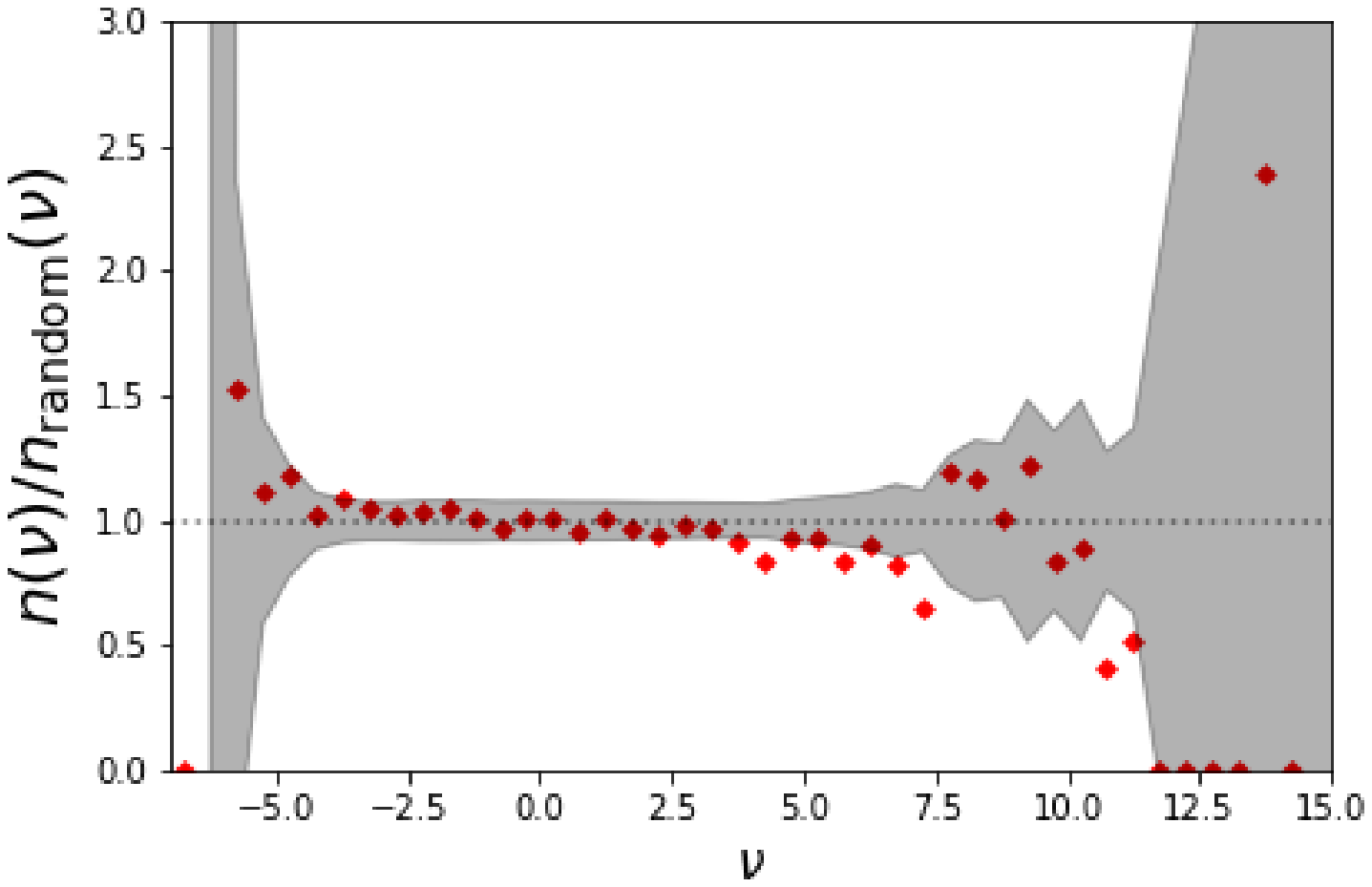}}
\caption{Differences in peak statistics around the lowest convergence peak. 
Horizontal axis shows $S/N$ defined by equation~(\ref{eq.sn_peak}). 
Vertical axis gives the ratio around the number of peaks around it to that at the random points.
The source redshift is $z_s=0.713$.
The red points indicate the results measured at the underdense regions.
The shaded region depicts the standard deviation estimated from the 500 random points.
Each panel shows results for a given search radius $\theta_r$:
{\it Top left:} $\theta_r=5$ degrees, {\it Top right:} $\theta_r=10$ degrees, {\it Middle left:} $\theta_r=20$ degrees, {\it Middle right:} $\theta_r=30$ degrees and {\it Bottom:} $\theta_r=50$ degrees.}
\label{fig.conv_void}
\end{figure*}

Figure~\ref{fig.conv_void} shows the ratios of the number of peaks at the lowest peak in the convergence maps to those for the 500 random points for different search radii $\theta_r=5, 10, 20, 30$ and $50$ degree, respectively.  
The red points show the ratios of the number counts toward the underdense region to the average number counts for the random points.
The shaded regions indicate $1\sigma$ obtained from the random points.
We found a decreasing number count trend at large $S/N$ regions.
As described in the previous subsection, structures grow slower in low density regions.
As these massive haloes are tightly correlated with high $S/N$ peaks in convergence maps \citep{2004MNRAS.350..893H}, the decrement of the haloes can reduce the number of high peaks.
On the other hand, the number of peaks in the low $S/N$ region increases because of lack of matter in the line of sight.
Thus, we calculated $S/N$ described in equation~(\ref{eq.sn}) to evaluate the significance of the differences in the peaks.  
In the estimate, we excluded bins whose number count in the random points is zero.  
Table~\ref{tab.sn} shows the result for different search radii.
We found that the total environmental effects on peak statistics would be detected at $S/N\geq5$.

\begin{table}
\caption{Singnal-to-Noise ratios for peak statistics at the lowest peak for each search radius $\theta_r$. Column~(1): radius used for a peak statistics. Column~(2): total $S/N$ estimated with equation~(\ref{eq.sn}). Column~(3): number of bins.}
\begin{center}
\begin{tabular}{|c|c|c|}
radius $\theta_r$ [degree] & S/N  & number of bins\\ \hline\hline
5                                        & 3.53 & 46\\	
10       				& 4.48 & 52\\
20					& 3.87 & 53\\
30					& 5.27& 53\\
50					& 4.96 & 54
\end{tabular}
\end{center}
\label{tab.sn}
\end{table}

\begin{figure*}
\subfigure{\includegraphics[width=1.0\columnwidth]{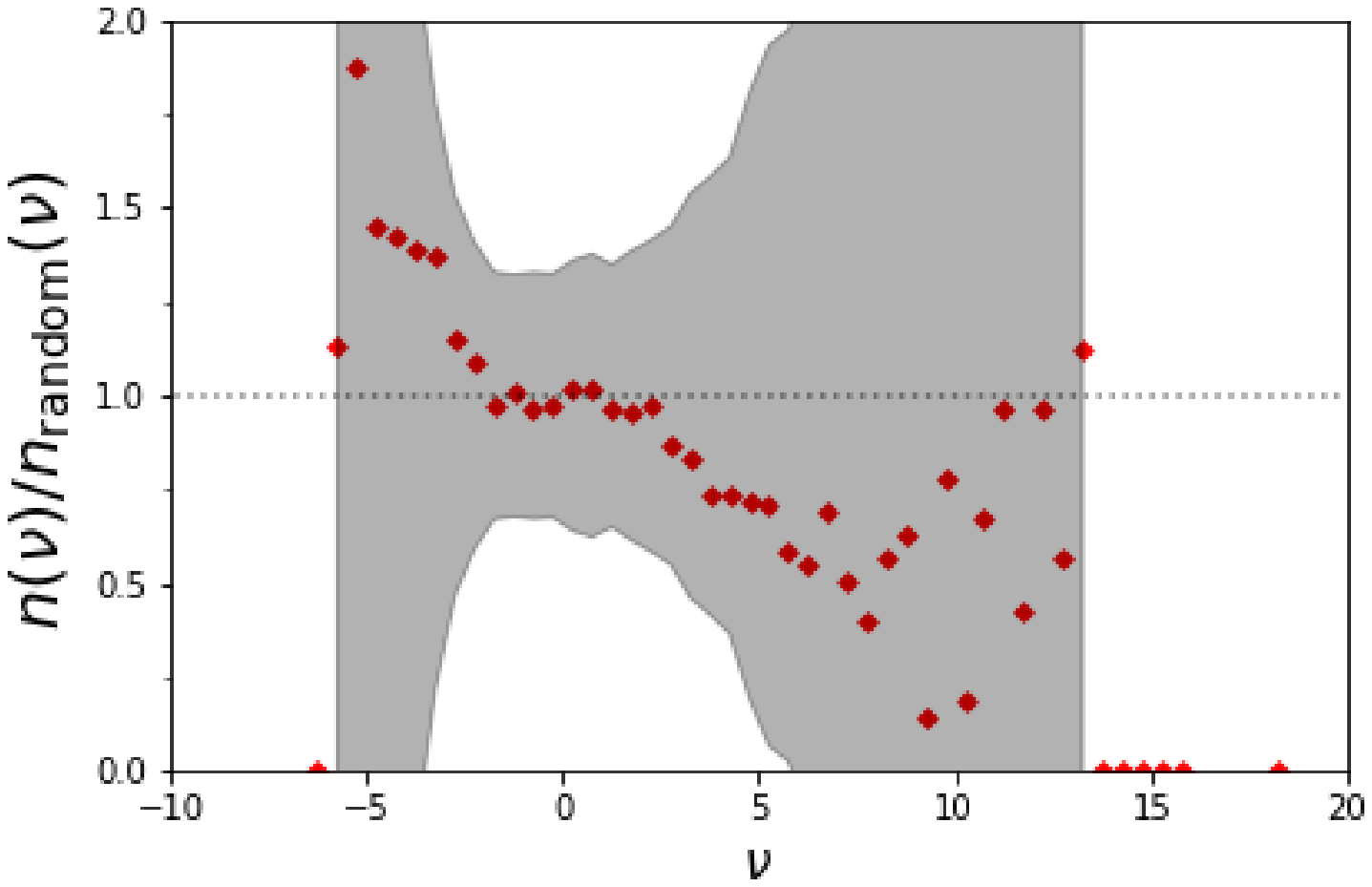}}
\subfigure{\includegraphics[width=1.0\columnwidth]{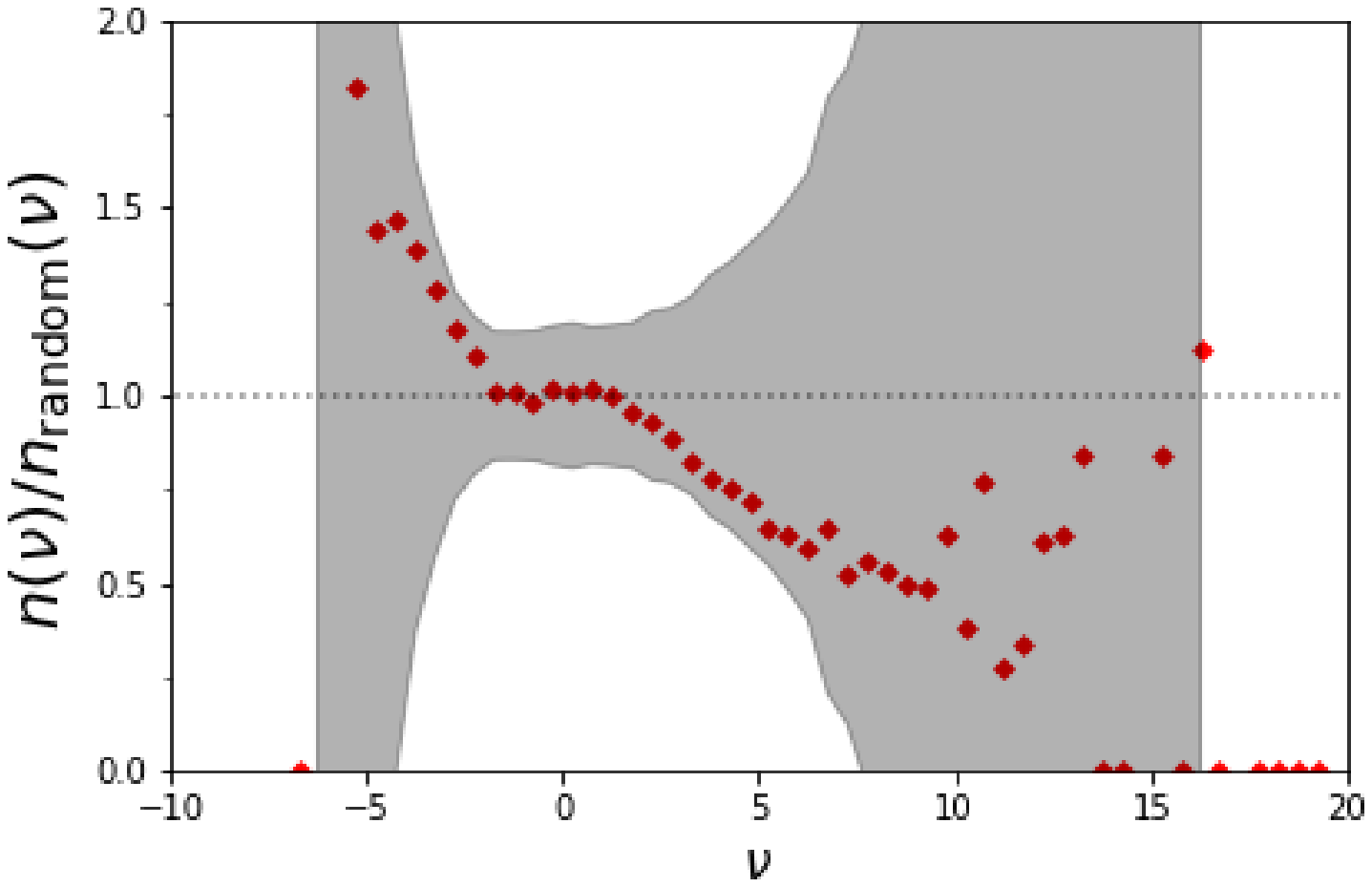}}
\subfigure{\includegraphics[width=1.0\columnwidth]{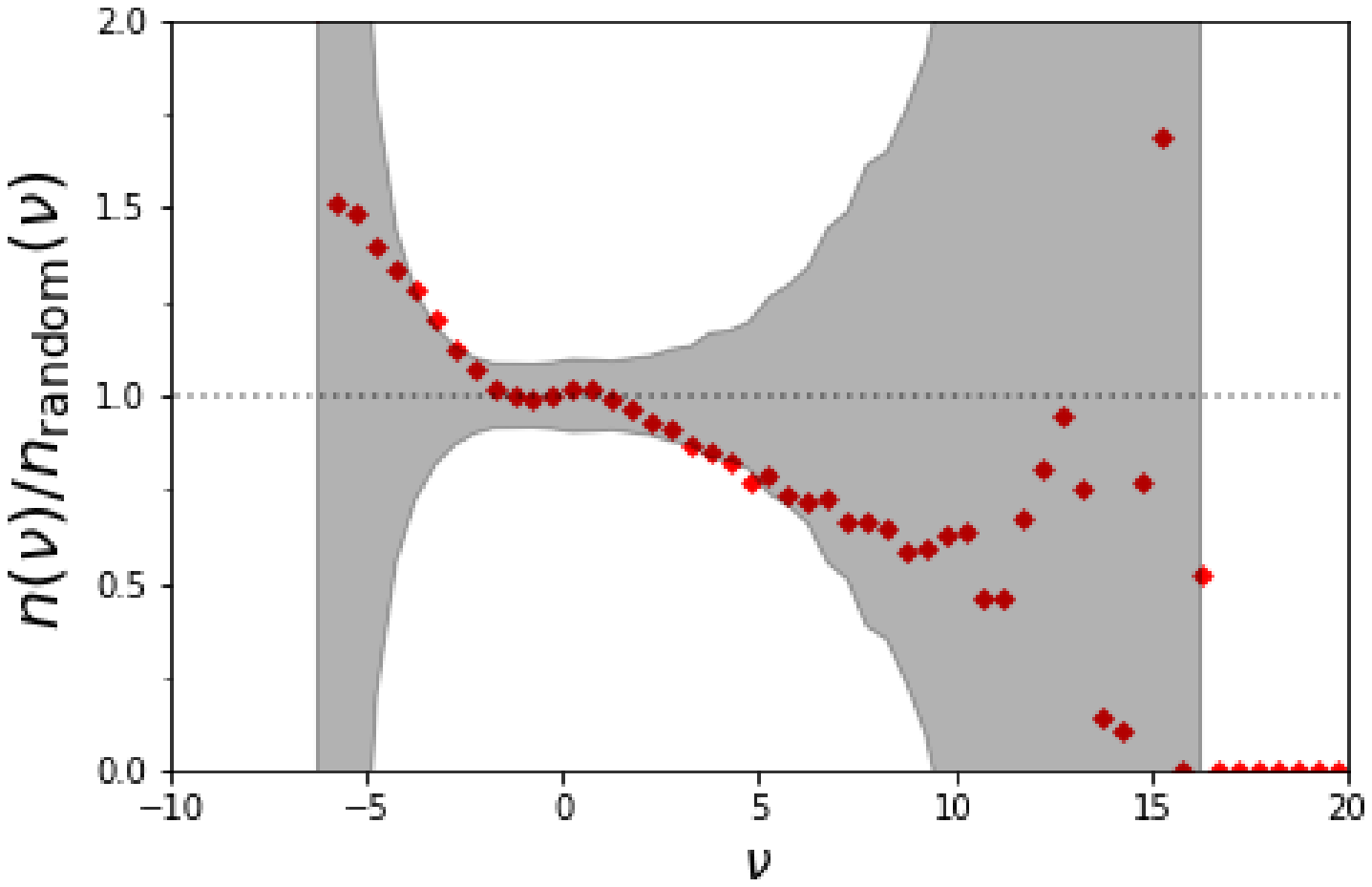}}
\subfigure{\includegraphics[width=1.0\columnwidth]{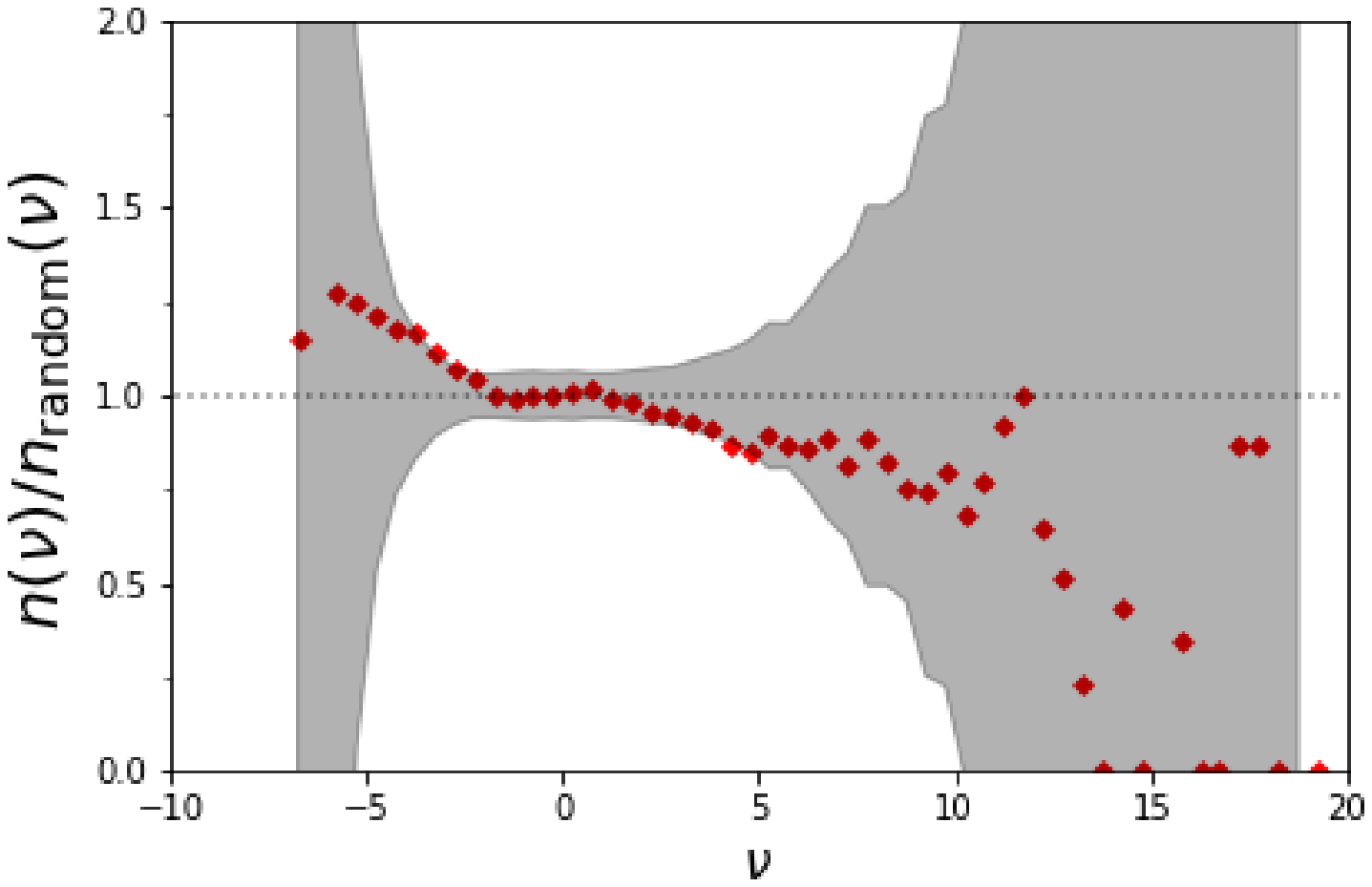}}
\subfigure{\includegraphics[width=1.0\columnwidth]{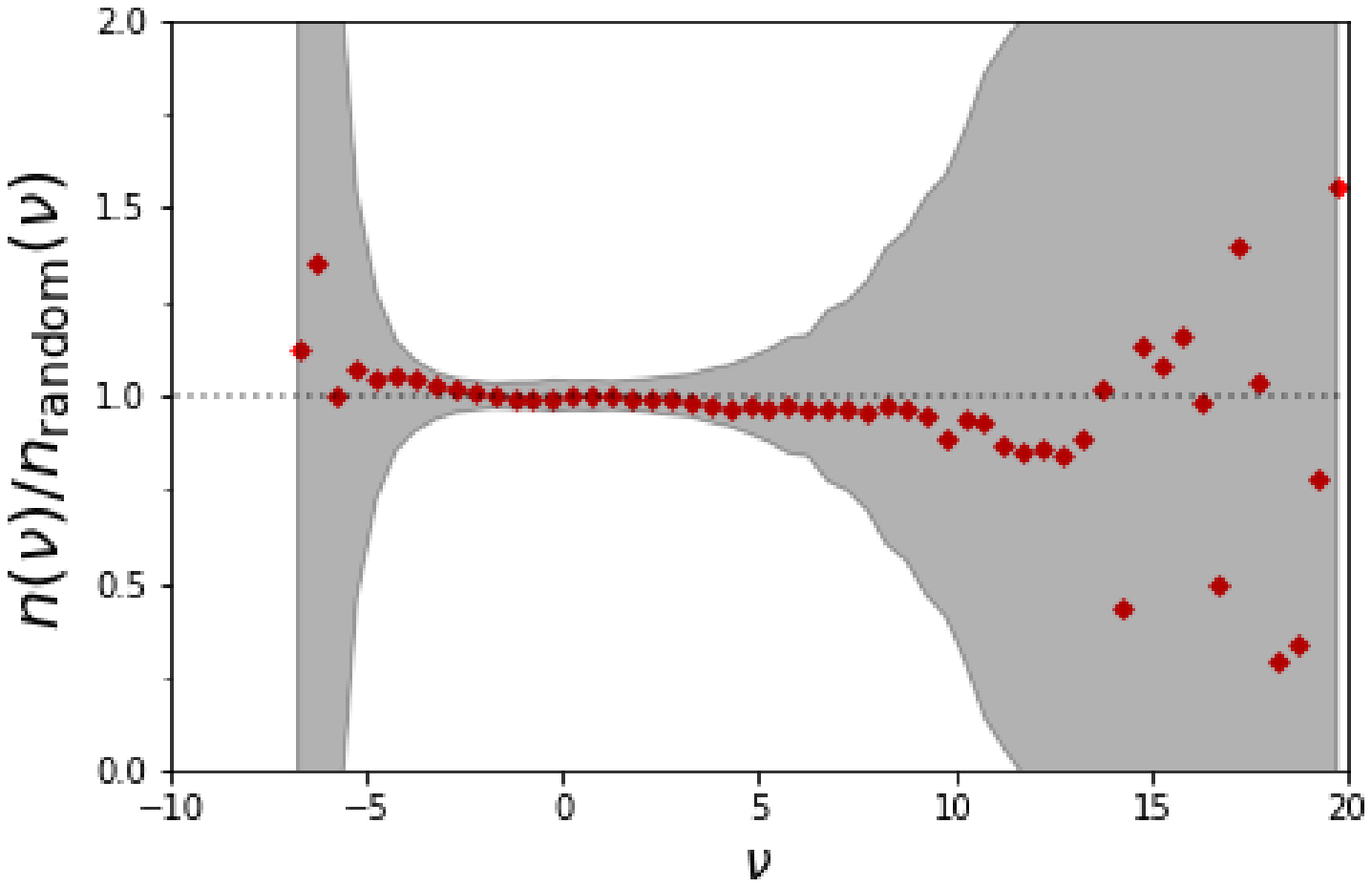}}
\caption{Similar to figure~\ref{fig.conv_void}, but results are obtained by the stacking method.  }
\label{fig.stack_void}
\end{figure*}

Moreover, we investigated the average property of peak statistics around the large underdense regions by stacking the number toward the underdense regions.
Figure~\ref{fig.stack_void} and table~\ref{tab.stacksn} show the ratios and $S/N$ for stacking analysis with the 106 low convergence regions. 
A similar trend for the lowest underdense region is observed, indicating the necessity of considering the environmental effects of large-scale structures on peak statistics.
Compared with the signal-to-noise ratio for the halo number count, the significances for the peak statistics are approximately two times lager
as the peak statistics trace not only positive peaks mostly corresponding to massive haloes but also negative peaks as well.
Therefore, large underdense regions affect both positive and negative peaks.

\begin{table}
\caption{Similar to table~\ref{tab.sn}, but the result are for the stacking analysis.}
\begin{center}
\begin{tabular}{|c|c|c|}
radius $\theta_r$ [degree] & S/N  & number of bins\\ \hline\hline
5                                        & 0.92 & 46\\	
10       				& 1.47 & 52\\
20					& 2.14 & 54\\
30					& 2.31 & 53\\
50					& 1.07 & 54
\end{tabular}
\end{center}
\label{tab.stacksn}
\end{table}

\section{Conclusions}
\label{sec.con}
We have studied the environmental effects of large underdense regions on halo number counts and weak lensing peaks in convergence maps. 
To observe the properties, we have used all-sky simulations in the {\it WMAP} 9yr cosmology.
We have smoothed the convergence maps with a smoothing scale of $20$ degrees and selected 106 underdense regions by adopting a threshold to the low peaks in the smoothed maps.
We have measured the number counts of the haloes at the lowest convergence peak and the average values in the large underdense regions through the stacking method.   
We have found that the number counts for massive haloes decreases and their significances with respect to the average cosmic values are $S/N\geq3$ in the total mass bin.
Moreover, we have measured the differences in the weak lensing peak counts in the same manner as the halo number counts.
We found that the number of positive peaks around the underdense regions decreases whereas that of the negative peaks increases.
This is because the decrement of massive haloes corresponding to the high peaks reduces the number count of positive peaks and the large underdense regions along the line of sight enhance the number counts of the negative peaks.
The significance in the differences in the peak count are $S/N\sim5$ in total for the largest case, indicating that the environmental effects on growth of haloes around large underdense regions affect weak lensing analysis of peak statistics. 
Therefore, we would be able to confirm the presence of supervoids toward the Cold Spot through measurement of the statistical difference in weak lensing observations. 

We further investigated the differences in the number count of dark matter haloes in the redshift range of $0\leq z\leq0.6$ and measured only on the sky.
Decrement of the halo number count becomes more significant for reduced redshift range of haloes, which indicates that massive haloes analysed in this study can be observed as galaxies and galaxy clusters.
Future multi-fibre spectroscopic surveys such as {\it Subaru/Prime Focus survey} \citep{2012SPIE.8446E..0YS} will be able to give a tighter constraint on the presence of the populated supervoids through the environmental effects.

\section*{Acknowledgements}
We thank an anonymous referee for giving useful comments and improving the manuscript. 
We thank R. Takahashi, T. Hamana and M. Shirasaki for carrying out the simulations.
We would like to thank K.Umetsu, T. Okumura and Y.Toba for useful comments and discussions.
This work is supported in part by the Ministry of Science and Technology of Taiwan (grant MOST 106-2628-M-001-003-MY3) and by Academia Sinica (grant AS-IA-107-M01).
This work was supported by NAOJ ALMA Scientific Research Grant Number 2018-07A. 
The second author was in part supported by JSPS KAKENHI Grant Number JP 17H02868
Numerical computations presented in this paper were in part carried out on the general-purpose PC farm at Center for Computational Astrophysics, CfCA, of National Astronomical Observatory of Japan.
Data analyses were (in part) carried out on common use data analysis computer system at the Astronomy Data Center, ADC, of the National Astronomical Observatory of Japan.





\bibliographystyle{mnras}
\bibliography{mn-jour,bibtex}



\appendix

\section{Transfer function}
\label{ap:trans}
In this section, we will describe the transfer function derived in \citet{1998ApJ...496..605E}.
The transfer function is written as a sum of the baryon and cold dark matter contributions, and described with a matter density parameter $\Omega_{\rm m} = \Omega_{\rm CDM} + \Omega_{\rm b}$ as
\begin{equation}
T(k)= \frac{\Omega_{\rm b}}{\Omega_{\rm m}}T_{\rm b}(k)+\frac{\Omega_{\rm CDM}}{\Omega_{\rm m}}T_{\rm CDM}(k),
\label{eq.transfer}
\end{equation}
where $\Omega_{\rm b}$ and $\Omega_{\rm CDM}$ are the density parameters for baryon and cold dark matter, respectively.
The transfer function of the CDM term has to take into account the suppression effect from baryon and is described as
\begin{equation}
T_{\rm CDM}(k)=f\tilde{T}_0(k, 1, \beta_c) + (1-f)\tilde{T}_0(k, \alpha_c, \beta_c).
\end{equation}
\begin{equation}
f=\frac{1}{1+(ks/5.4)^4},
\end{equation}
where $s$ is the sound horizon defined as
\begin{equation}
s=\frac{44.5 {\rm ln}(9.83/\Omega_{\rm m}h^2) }{\sqrt{1+10(\Omega_{\rm b}h^2)^{3/4}}}.
\end{equation}
$h$ is the Hubble parameter defined as
\begin{equation}
h={\rm H}_0/(100 {\rm km~s^{-1}~Mpc^{-1}} ).
\end{equation}
\begin{equation}
\tilde{T}_0(k, \alpha_c, \beta_c)=\frac{ {\rm ln}(e+1.8\beta_cq) }{ {\rm ln}(e+1.8\beta_cq)+Cq^2 },
\end{equation}
\begin{equation}
C=\frac{14.2}{\alpha_c}+\frac{386}{1+69.9q^{1.08}}
\end{equation}
The variables are described as
\begin{equation}
q=\frac{k}{13.41k_{\rm eq}},
\end{equation}
\begin{equation}
\alpha_c=a_1^{-\Omega_{\rm b}/\Omega_{\rm m}}a_2^{(\Omega_{\rm b}/\Omega_{\rm m})^3},
\end{equation}
\begin{equation}
a_1=(46.9\Omega_{\rm m}h^2)^{0.67}\left[1+(32.1\Omega_{\rm m}h^2)^{-0.532}\right],
\end{equation}
\begin{equation}
a_2=(12\Omega_{\rm m}h^2)^{0.424}\left[1+(45\Omega_{\rm m}h^2)^{-0.582}\right],
\end{equation}
\begin{equation}
\beta_c^{-1}=1+b_1[(\Omega_{\rm CDM}/\Omega_{\rm m})^{b_2}-1],
\end{equation}
\begin{equation}
b_1=0.944[1+(458\Omega_{\rm m}h^2)^{-0.708}]^{-1},
\end{equation}
\begin{equation}
b_2=(0.395\Omega_{\rm m}h^2)^{-0.0266}.
\label{eq.b2}
\end{equation}
$k_{\rm eq}$ is the particle horizon scale.
The term for baryon can be described as
\begin{equation}
T_{\rm b}(k)=\left[ \frac{\tilde{T}_0(k, 1, 1)}{1+(ks/5.2)^2}+\frac{\alpha_{\rm b}}{1+(\beta_{\rm b}/ks)^3}e^{-(k/k_{\rm silk})^{1.4}} \right] j_0(k\tilde{s}),
\end{equation}
where $j_0$ is the spherical Bessel function and $k_{\rm silk}$ is the Silk scale.
$\alpha_{\rm b}$ is the parameter for characterizing amplitude of suppression.
$\beta_{\rm b}$ and $\tilde{s}$ are defined as
\begin{equation}
\beta_{\rm b}=0.5+\frac{\Omega_{\rm b}}{\Omega_{\rm m}}+\left(3-2\frac{\Omega_{\rm b}}{\Omega_{\rm m}}\right)\sqrt{(17.2\Omega_{\rm m}h^2)^2+1},
\end{equation}
\begin{equation}
\tilde{s}(k)=\frac{s}{[1+(\beta_{\rm node}/ks)^3]^{1/3}},
\end{equation}
where $\beta_{\rm node}$ is described as
\begin{equation}
\beta_{\rm node}=8.41(\Omega_{\rm m}h^2)^{0.435}.
\label{eq.betanode}
\end{equation}



\bsp	
\label{lastpage}
\end{document}